\documentclass[prb,twocolumn,floatfix,showpacs]{revtex4-1}
\usepackage{amsmath,amssymb,graphicx,psfrag}
\usepackage{multirow}
\usepackage{color}
\def\bea{\begin{eqnarray}}
\def\eea{\end{eqnarray}}
\newcommand{\rv}{\vec{r}}

\newcommand{\kv}{{\vec{K}}}

\newcommand{\uj}{u^{(j)}}
\newcommand{\ujs}{u^{(j)*}}

\newcommand{\khj}{\hat{k}^{(j)}}
\newcommand{\Rv}{\vec{R}}

\newcommand{\Aj}{A^{(j)}}
\newcommand{\Ajs}{A^{(j)*}}
\newcommand{\epsnew}{\tilde{\epsilon}}

\begin{document}

\title{Phase-field-crystal study of grain boundary premelting and shearing in bcc iron}

\author{Ari Adland}
\affiliation{Department of Physics and Center for Interdisciplinary Research
on Complex Systems, Northeastern University, Boston, MA 02115 USA}
\author{Robert Spatschek}
\affiliation{Max-Planck-Institut f\"ur Eisenforschung GmbH, D-40237 D\"usseldorf, Germany}
\author{Dorel Buta}
\author{Mark Asta}
\affiliation{Department of Materials Science and Engineering, University of California, Berkeley, California, 94720 USA}
\author{Alain Karma}
\affiliation{Department of Physics and Center for Interdisciplinary Research
on Complex Systems, Northeastern University, Boston, MA 02115 USA}
%\thanks{Corresponding author: a.karma@neu.edu}

\pacs{61.72.Mm, 64.70.D-, 81.30.Fb, 05.40.Ca}

\date{\today}

\begin{abstract}
We use the phase-field-crystal (PFC) method to investigate the equilibrium premelting and nonequilibrium shearing behaviors of $[001]$ symmetric tilt grain boundaries (GBs) at high homologous temperature over the complete range of misorientation $0<\theta<90^\circ$ in classical models of bcc Fe. We characterize the dependence of the premelted layer width $W$ as a function of temperature and misorientation and compute the thermodynamic disjoining potential whose derivative with respect to $W$ represents the structural force between crystal-melt interfaces due to the spatial overlap of density waves. The disjoining potential is also computed by molecular dynamics (MD) simulations, for quantitative comparison with PFC simulations, and coarse-grained amplitude equations (AE) derived from PFC that provide additional analytical insights. We find that, for GBs over an intermediate range of misorientation ($\theta_{\rm min}<\theta<\theta_{\rm max}$), $W$ diverges as the melting temperature is approached from below, corresponding to a purely repulsive disjoining potential, while for GBs outside this range ($\theta<\theta_{\rm min}$ or $\theta_{\rm max}<\theta<90^\circ$), $W$ remains finite at the melting point, with its value corresponding to a shallow attractive minimum of the disjoining potential. The misorientation range where $W$ diverges predicted by PFC simulations is much smaller than the range predicted by MD simulations when the small dimensionless parameter $\epsilon$ of the PFC model, where $\epsilon^{-1/2}$ is proportional to the ratio of the solid-liquid interface width to the lattice spacing, is matched to liquid structure factor properties. However it agrees well with MD simulations with a lower $\epsilon$ value chosen to match the ratio of bulk modulus and solid-liquid interfacial free-energy, consistent with the amplitude-equation prediction that $\theta_{\rm min}$ and $90^\circ-\theta_{\rm max}$ scale as $\sim \epsilon^{1/2}$. The incorporation of thermal fluctuations in PFC is found to have a negligible effect on this range. In response to a shear stress parallel to the GB plane, GBs in PFC simulations exhibit coupled motion normal to this plane, with a discontinuous change of the coupling factor as a function of $\theta$ that reflects a transition between two coupling modes, and/or sliding (shearing of the two grains). Partial sliding is only observed over a range of misorientation that is a strongly increasing function of temperature for $T/T_M\ge 0.8$ and matches roughly the range where $W$ diverges at the melting point. The coupling factor for the two coupling modes is in excellent quantitative agreement with previous theoretical predictions [J. W. Cahn, Y. Mishin, and A. Suzuki, Acta Mater. 54, 4953 (2006)]. 
\end{abstract}

\maketitle 

%%%%%%%%%%%%%%%%%%%%%%%%%%%%%%%%%%%%%%%%%%%%%%%%%%%%%%%%%%%
\section{Introduction}

Grain boundaries (GBs) strongly influence the mechanical behavior and other materials properties. For this reason, they have been widely studied both experimentally \cite{SuttonBalluffi1995} and computationally \cite{Mishinetal2010} for decades. At high homologous temperature, GBs can display pronounced disorder, manifested in the most extreme case by the formation of nanometer-scale intergranular films with liquid-like properties. The formation of those films below the bulk melting point, typically referred to as GB premelting, can dramatically reduce shear resistance and lead to catastrophic materials failure. This phenomenon is of interest for predicting the formation of solidification defects associated with the formation of those intergranular films, which can lead to hot cracking during the late stages of solidification \cite{Rappazetal2003,Wangetal2004,Astaetal2009}, and more generally for understanding the microstructure and mechanical behavior of structural alloys at high homologous temperature.

GB premelting has been widely studied experimentally 
\cite{Chanetal1985,BallufiMaurer1988,HsiehBalluffi1989,Masumuraetal1972,Voldetal1972,Watanabeetal1984,Dashetal1995,Inokoetal1997,Divinskietal2005,Luoetal2005,Guptaetal2007}
as well as theoretically. Theoretical approaches include discrete lattice models \cite{KikuchiCahn1980} and molecular dynamics (MD) simulations \cite{BroughtonGilmer1986,vonAlfthan2007,WilliamsMishin2009,Hoytetal2009,Fensinetal2010,Olmstedetal2011}, as well as conventional phase-field models \cite{LobkovskyWarren2002,Tangetal2006,Mishinetal2009,Wangetal2008}, which either exploit an orientational order parameter \cite{LobkovskyWarren2002,Tangetal2006} or multiple phase-fields \cite{Mishinetal2009,Wangetal2008} to distinguish between grains, and the phase-field-crystal (PFC) method \cite{Berryetal2008,Mellenthinetal2008}, which resolves the crystal density field on an atomic scale and hence naturally models crystal defects such as isolated dislocations and GBs. 

Two central issues in GB premelting have been the determination of the premelted layer width $W$ and the quantification of the fundamental forces that control this width. It has been generally difficult to address those issues experimentally due to the challenges
inherent in observing internal materials
interfaces such as GBs. Limited observations to date support the existence of a nanometer-thick premelted layer in pure materials a few degrees below the bulk melting point and there is more ample evidence for premelting in alloys. More definitive insights into those issues has been provided by recent PFC \cite{Mellenthinetal2008} and MD simulation \cite{Hoytetal2009,Fensinetal2010} studies that have characterized quantitatively the structural forces underlying GB premelting. 
Quantification of those forces has been obtained by computing the ``disjoining potential'' $V(W)$ defined through the   
excess Gibbs free-energy per unit of grain boundary area  
\begin{equation}
G_{\rm exc}(W,T)=\Delta G(T)W+2\gamma_{sl}+V(W),
\label{gbexc}
\end{equation}
where $\Delta G=G_s-G_l$ is the bulk Gibbs free-energy difference between liquid ($G_l(T)$) and solid ($G_s(T)$) and $\gamma_{sl}$ is the solid-liquid interfacial free-energy. With this definition, $V(W)$ represents the part of this excess due to the overlap of crystal density waves from the two grains on each side of the GB. Hence the derivative $-dV(W)/dW$ measures the force between crystal-melt interfaces due to this overlap, which can be either repulsive or attractive depending on whether the sign of $-dV(W)/dW$ is positive or negative, respectively. It is worth noting that the disjoining potential can be defined in different thermodynamic ensembles. While Eq.~(\ref{gbexc}) defines it in the Gibbs ensemble for MD simulations, we shall also use the grand canonical and canonical ensembles for PFC and amplitude equation simulations, respectively.   

In the simplest formulation of the disjoining potential, $W$ can be viewed as the width of a liquid layer sandwiched between two atomically sharp solid-liquid phase boundaries \cite{Rappazetal2003}. Furthermore, $V(W)$ is assumed to have a simple exponentially decaying form (see e.g.~Ref.~\onlinecite{Widom1978})
\begin{equation}
V(W)=\Delta\gamma \exp(-W/\delta),\label{simpleVofW}
\end{equation}
where $\Delta\gamma=\gamma_{gb}-2\gamma_{sl}$ is the difference between the GB energy ($\gamma_{gb}$) and the excess free-energy of two separated solid-liquid interfaces ($2\gamma_{sl}$).  Substitution of this form into Eq. (\ref{gbexc}) and minimization of $G_{\rm exc}(W,T)$ with respect to $W$  predicts that, for $\Delta\gamma>0$, the GB is ``dry'' ($W=0$) up to a bridging temperature $T_b<T_M$ beyond which $W$ increases and diverges logarithmically at the melting temperature $T_M$, while 
for $\Delta\gamma<0$, the GB remains dry up to a superheated temperature $T^*>T_M$; for $T>T^*$, the dry  GB state becomes thermodynamically unstable and the solid melts completely.  

Recent PFC \cite{Mellenthinetal2008} and MD studies \cite{Hoytetal2009,Fensinetal2010} have shed light on several important aspects of the disjoining potential and GB premelting that are not predicted by this simple model. First, while $V(W)$ is indeed found to be purely repulsive for some high-energy GBs, with a form that can be approximately fitted by Eq.~(\ref{simpleVofW}) \cite{Mellenthinetal2008,Hoytetal2009} it is not purely attractive (i.e. $dV(W)/dW>0$ for all $W$) for lower energy GBs. For the latter GBs, $V(W)$ exhibits a long-range attractive part, due to the overlap of the decaying tails of crystal density waves from each grain, as qualitatively predicted by Eq. (\ref{simpleVofW}), but also a short-range repulsive part associated with the formation energy of crystal defects (e.g., dislocations for low-angle GBs) when $W$ becomes comparable to the lattice spacing \cite{Mellenthinetal2008}. As a result, for GBs where $W$ does not diverge at $T_M$, $V(W)$ generically exhibits a minimum, which implies that the premelted layer width remains finite at the melting point.  Importantly, this nanometer-scale width represents a compromise between the long-range-attractive and short-range-repulsive parts of the disjoining potential that are both structural in nature, since they involve the decaying tails of the crystal density field and crystal-defect formation energy, respectively. 

More generally, $V(W)$ also contains an attractive part due to London dispersion forces that are not accounted for in PFC and MD simulations, but play an important role in other systems such as ceramic materials \cite{Clarke1987}. However, in metallic systems, dispersion forces can be estimated to only contribute an attractive tail to $V(W)$ whose magnitude is less than a mJ/m$^2$ for $W$ on the nanometer scale. In contrast, MD computations of  $V(W)$ in pure Ni \cite{Hoytetal2009,Fensinetal2010} show that $V(W)$ has a magnitude of tens of mJ/m$^2$ for $W$ in this same range (consistent with the prediction of Eq.~({\ref{simpleVofW}) in a purely repulsive case). Hence, GB premelting appears clearly dominated by structural forces in metallic systems, as further supported by the present study in pure Fe.   

Recent PFC \cite{Mellenthinetal2008} and MD \cite{Olmstedetal2011} studies have also shed light on the condition under which a GB will be ``wet'' (with a diverging $W$) or ``dry'' (with a finite $W$) at the melting point (neglecting dispersion forces). The classic wetting condition $\Delta\gamma=\gamma_{gb}-2\gamma_{sl}>0$ turns out to give a grossly inaccurate prediction when a low-temperature GB energy is used to compute $\gamma_{gb}$. This inaccuracy originates from the fact that the GB energy at the melting point $\gamma_{gb}(T_M$) is generally significantly smaller than at low temperature, with a large part of the decrease due to the elastic softening of the material (decrease of the elastic constants) at high homologous temperature \cite{Mellenthinetal2008}. Hence, a more accurate prediction of when a GB will be wet or dry has been obtained in PFC simulations by using the wetting condition $\Delta\gamma>0$ in conjunction with a value of $\gamma_{gb}(T_M)$ that takes into account this elastic softening \cite{Mellenthinetal2008}. The extension of this approach to MD simulations of Fe for symmetric [001] tilt GBs has yielded reasonably good predictions of which GBs will be fully wet or dry over a complete range of misorientation \cite{Olmstedetal2011}. This study also associated departures from this prediction to the existence of a dislocation-pairing transition, which provides an additional means for the GB to lower its free-energy, distinct from elastic softening. 

The formation of an integranular liquid-like layer would be expected physically to lead to dramatic decrease of shear resistance of a wet GB, and hence relative sliding of the two grains when a shear stress is applied parallel to the GB plane. In contrast, a low angle dry GB with intact solid bridges in between isolated dislocations with premelted cores would be expected to support shear more like a static solid. However, both theoretical \cite{CahnTaylor2004,Cahnetal2006,IvanovMishin2008,TrauttMishin2012,Trauttetal2012} and experimental 
\cite{Gorkayaetal2009,Mompiouetal2009,Molodovetal2011} studies over the last decade have shown that a dry GB generically moves normal to itself under an applied shear stress. This motion coupled to shear, commonly referred to as ``coupled motion'', is characterized by a relationship $v_\parallel=\beta v_n$ between the translation velocity of the two grains parallel to the GB plane, $v_\parallel$, and the GB velocity normal to this plane, $v_n$.  For symmetric tilt GBs,  analytical predictions for the coupling factor $\beta(\theta)$ (where $\theta$ is the misorientation) based on geometrical arguments have been validated by both MD simulations \cite{Cahnetal2006} and experiments \cite{Gorkayaetal2009,Mompiouetal2009}. 

An important question for mechanical behavior at high homologous temperature is when will a GB exhibit coupled or sliding motion. An MD study of symmetric tilt GBs in pure Ni \cite{Cahnetal2006} showed that the range of misorientation where sliding is observed increases close to the melting point, consistent with the view that  the formation of an intergranular liquid-like film favors sliding over coupling. However, a recent combined PFC and MD study \cite{Trauttetal2012} also showed that asymmetrical GBs can exhibit sliding unrelated to premelting, so that the transition from coupling to sliding is generally more complex. 

In this paper, we investigate GB premelting and shearing and their relationship using the PFC method \cite{Elderetal2002,ElderGran2004,Stefanovicetal2006,Elderetal2007,WuKarma2007,Wuetal2010,Jaatinenetal2010}. 
We use the simplest PFC model \cite{Elderetal2002,ElderGran2004} with the same free-energy function as the Swift-Hohenberg model of pattern formation \cite{SwiftHohenberg1977}, which favors hexagonal  and bcc ordering in two and three dimensions (2D and 3D), respectively. This model can be interpreted \cite{Elderetal2007} as a considerably simplified version of classical density function theory (DFT) \cite{HaymetOxtoby1981,Lairdetal1987,Singh1991,HarwellOxtoby1984,ShenOxtoby1996a,ShenOxtoby1996b} where the crystal density field is dominated by the set of primary reciprocal lattice vectors. 
With suitable choices of parameters for Fe (see Ref.~\onlinecite{WuKarma2007}), this model has been shown to
predict reasonable values of solid-liquid interfacial energy \cite{Wuetal2006,WuKarma2007} and GB energies
\citep{Jaatinenetal2010}, despite the uncontrolled truncation of many other sets of reciprocal lattice vectors. It has also proven capable of reproducing a subtle dislocation-pairing GB structural transition 
at high homologous temperatures \citep{Olmstedetal2011}, previously evidenced in a 2D PFC study of GB premelting \cite{Mellenthinetal2008}, and also observed in MD simulations \citep{Olmstedetal2011}. However, the only two PFC studies of GB premelting to date have remained qualitative \cite{Berryetal2008,Mellenthinetal2008}. The study of Berry {\it et al.} \cite{Berryetal2008} in 3D did not compute the disjoining potential and was not carried out for parameters of Fe. The one of Mellenthin {\it et al.} \cite{Mellenthinetal2008} computed this potential, revealing the existence of a shallow minimum for dry boundaries, but was limited to 2D hexagonal ordering. In addition, PFC studies of GB shearing have only been carried out recently for 2D square ordering \cite{Trauttetal2012}. 

Here we compute quantitatively the disjoining potential for $[001]$ symmetric tilt grain boundaries over the complete range of misorientation $0<\theta<90^\circ$ and also study quantitatively the response to an applied shear stress, distinguishing between regimes of coupling and sliding as a function of $\theta$ and temperature. We also investigate the role of thermal fluctuations on the disjoining potential by carrying out PFC simulations without and with the addition of Langevin noise that is uncorrelated in space and time. Following the standard approach, the magnitude of this noise is fixed by the fluctuation-dissipation theorem and we also introduce a short-wavelength cut-off of this noise, which is shown to be necessary to avoid unphysical divergences.  
To benchmark our results, we compare the disjoining potentials computed by PFC and MD simulations. The MD simulations are carried out using the same EAM potential \cite{Mendelevetal2003} that was previously used to calibrate PFC model parameters for Fe \cite{Wuetal2006,WuKarma2007,Wuetal2010}. With appropriate rescaling of length and energy, the PFC model can be expressed in a form that involves a single small dimensionless parameter $\epsilon$. In the context where PFC is derived from DFT, $\epsilon$ is uniquely determined by liquid-structure factor properties
\cite{Wuetal2006,WuKarma2007,Wuetal2010}. More generally, decreasing this parameter makes the freezing transition more weakly first order and increases the width of the solid-liquid interface in units of lattice spacing. To gain additional insights into the role of this parameter, we also compute the disjoining potential using amplitude equations \cite{Goldenfeldetal2005,Athreyaetal2006,Athreyaetal2007,Shiwa2009,Goldenfedletal2009,ChanGoldenfeld2009,Elderetal2010,SpatschekKarma2010}. These equations can be formally derived from the PFC model in the small $\epsilon$ limit 
using similar multiscale expansions introduced previously to analyze continuum models of pattern formation \cite{NewellWhitehead1969,Segel1969,CrossHohenberg1993,Gunaratneetal1994,Graham1996,Graham1998a,MatsubaNozaki1998,Graham1998b}. While amplitude equations have recently been shown to break down for high angle GBs because of issues related to frame invariance \cite{SpatschekKarma2010}, they are asymptotically exact for small $\epsilon$ and low angle GBs. Hence, in the present context, they allow us to derive scaling laws for the range of misorientation over which GBs exhibit a diverging liquid-layer width. Here we report amplitude-equation results that pertain to the numerical computation of disjoining potentials and the derivation of this scaling law. In a companion paper to this one, we present the results of a more in depth analytical and numerical amplitude-equation study of disjoining potentials. In this paper, we also treat the case (analogous to the computation of $\gamma$ surfaces) where two crystals of the same orientation are translated from each other in the plane of the interface   \cite{ShiftedCrystalPaper}.

The rest of this paper is organized as follows. In section \ref{pfc::sec}, we write down the PFC model and motivate the necessity of introducing a short-wavelength cut-off in the Langevin noise with a short calculation of the renormalization of the free-energy of the liquid state. In section \ref{gl::sec}, we write down the amplitude equations and relations that connect them to the PFC model. In section \ref{methodology::sec}, we then discuss the methodology to compute the disjoining potential with the PFC model without and with fluctuations and in MD simulations. The same methodology developed in Refs.~\onlinecite{Hoytetal2009,Fensinetal2010}, which relates the disjoining potential $V(W)$ to the probability density $P(W,T)$ of width fluctuations at different temperatures close to the melting point is used for both MD and PFC simulations with noise. PFC simulations without noise follow the same methodology developed in Ref.~\onlinecite{Mellenthinetal2008} using the grand canonical ensemble appropriate for PFC. In section \ref{gbpremelting::sec}, we then present the numerical results for disjoining potentials computed with the PFC approach without and with fluctuations, MD simulations, and amplitude equations, and we derive a scaling law for the critical angle below (or above) which a GB is dry or wet. Next, in section \ref{shearing::sec}, we present the PFC results for GB shearing. Our results are then summarized in section \ref{summary::sec}.

%%%%%%%%%%%%%%%%%%%%%%%%%%%%%%%%%%%%%%%%%%%%%%%%%%%%%%%%%%%
\section{PFC model}
\label{pfc::sec}

%%%%%%%%%%%%%%%%%%%%%%%%%%%%%%%%%%%%%%%%%%%%%%%%%%%%%%%%%%%
\subsection{Basic equations}

We use the simplest version of the PFC model \cite{Elderetal2002,ElderGran2004} with the same free-energy functional 
as the Swift-Hohenberg model of pattern formation \cite{SwiftHohenberg1977}, which is defined by
\begin{equation}
{\cal{F}}=\int d\vec{r}\left(\frac{\phi}{2}[a+\lambda(q_0^2+\nabla^2)^2]\phi+g\frac{\phi^4}{4}\right),
\end{equation}
where $\phi$ is a dimensionless measure of the crystal density field.  This free-energy favors  bcc ordering in 3D and has been shown to predict reasonably well the solid-liquid interfacial free-energy and its anisotropy as compared to MD simulations for parameters of Fe \cite{WuKarma2007}.

In order to minimize the number of relevant parameters, we introduce the scalings
\begin{eqnarray}
\label{epsconv}
\epsilon=-\frac{a}{\lambda q_0^4},\\
\vec{r'}=q_0\vec{r},\\
\psi=\sqrt{\frac{g}{\lambda q_0^4}}\phi, \label{pfc::eq1} \\
F=\frac{g}{\lambda^2 q_0^5}{\cal{F}},
\end{eqnarray}
leading to the dimensionless free-energy functional,
\begin{equation}
F=\int d\vec{r} \left( \frac{\psi}{2}[-\epsilon+(\nabla^2+1)^2]\psi+\frac{\psi^4}{4}\right),
\label{ffunc}
\end{equation}
where we have dropped the prime on $\vec{r}'$, so that $\vec{r}$ denotes the dimensionless position vector hereafter. 

We use a non-conserved dynamics of the form
\begin{equation}
\frac{\partial \psi}{\partial t}=-\frac{\delta \Omega}{\delta\psi}+\eta,
\label{fulldyn}
\end{equation}
where $\Omega=F-\mu\int d\vec{r}\psi$ is the grand potential, $\mu$ is the constant chemical potential, and $\eta$ is a Langevin 
noise that is uncorrelated in space and time with zero mean 
\begin{equation}
\langle \eta(\vec{r},t)\rangle=0,
\end{equation}
and variance
\begin{equation}
\langle \eta(\vec{r},t)\eta'(\vec{r}',t')\rangle= \Gamma \delta(\vec{r}-\vec{r}')\delta(t-t'),\label{noisedef}
\end{equation}
where
\begin{equation}
\Gamma = \frac{2k_BTg}{\lambda^2q_0^5},\label{Gammadef}
\end{equation}
measures the noise strength in our dimensionless PFC units. In simulations without noise, $\eta$ is set to $0$.  

While the present study could be equivalently carried out using the standard form of the PFC dynamics that conserves $\psi$ \cite{Elderetal2002,ElderGran2004}, we have found it useful to work in the grand canonical ensemble where $\mu$ plays an analogous role to temperature in the Gibbs ensemble in which the disjoining potential is usually defined via Eq.~(\ref{gbexc}). We can then use thermodynamic relations to relate results in those two ensembles. 

The higher order spatial derivatives make the PFC equations stiff in real space. We therefore conduct our simulations in spectral space, as originally presented in \cite{Mellenthinetal2008} with the details for the addition of noise presented in Appendix \ref{dynamicsA}.

To simulate GB shearing, we stay in the grand canonical ensemble, but use the modified 
PFC model with inertia (i.e. with an additional $\partial^2\psi/\partial t^2$ term)
\cite{Stefanovicetal2009}
\begin{equation}
\frac{\partial^2\psi}{\partial t^2}+\alpha\frac{\partial \psi}{\partial t} = -\frac{\delta F}{\delta \psi}+\mu,
\label{dotdyn}
\end{equation}
which has the advantage to quickly relax the elastic field via propagative phonon-like modes, as opposed to purely diffusive modes. This allows us to shear the crystal at velocities much faster than those we could use with the standard diffusive dynamics. Eq. (\ref{dotdyn}) is simulated with a modified spectral scheme presented in Appendix \ref{dynamicsB}.

%%%%%%%%%%%%%%%%%%%%%%%%%%%%%%%%%%%%%%
\subsection{Short-wavelength noise cutoff}
\label{cutoff}

In this section, we examine how to choose the short wavelength (high $k$) cutoff  of the Langevin noise in the PFC model. Previous studies of Langevin noise in PFC have used such a cutoff to carry out simulations \cite{Tothetal2011,Teeffelenetal2009,Granasyetal2011,Emmerishetal2012}. The issue of whether introducing noise may lead to a double counting of fluctuations has often been discussed. On the one hand, the free-energy functional of classical DFT, or its simplified version in PFC, already represents a mean-field average of atomic scale fluctuations. Thus adding back fluctuations would appear to be double counting. However, without the addition of noise, dynamical DFT or PFC simulations do not reproduce long-wavelength hydrodynamic fluctuations, e.g. capillary fluctuations of the solid-liquid interface \cite{Hoyetal2001}, which are present in a real system. Hence, in situations where those fluctuations, which are not double counted, are of interest, it makes sense to add noise with a short-wavelength cutoff to avoid double counting short-wavelength fluctuations already incorporated in the DFT or PFC free-energy functional.

In general, the magnitude of the noise is set by the fluctuation-dissipation theorem. However, the choice of the cutoff between not-double-counted long-wavelength and double-counted short-wavelength fluctuations contains some degree of arbitrariness. A common choice is to cut off the noise on the lattice scale, which is equivalent to choosing the maximum wavector of the noise $k_{max}$ of order unity in our dimensionless units where length is measured in unit of $q_0^{-1}$. To make this choice somewhat more precise, we compute here explicitly how the noise renormalizes the grand potential (Landau free-energy) of the liquid. This renormalized noise-induced excess potential is found to diverge $\sim k_{max}^3$. We then compute the value of $k_{max}$ at which this excess potential is equal to the barrier height of the double well potential between solid and liquid of the noiseless PFC. This value sets an upper bound on $k_{max}$, beyond which the noise is too strong and essentially destroys solid-liquid coexistence.

To compute the renormalized grand potential, we substitute $\psi =\bar{\psi}_l+\delta\psi$ into Eq. (\ref{ffunc}}) and expand to quadratic order 
the excess $\Delta \Omega\equiv \Omega_{(\eta\ne 0)}-\Omega(\bar{\psi}_l)$, which is the difference between the grand potential of a liquid with density fluctuations driven by the Langevin noise and a liquid of constant density $\bar{\psi}_l$. Since $\mu\int d\vec{r}\delta\psi=0$ gives no contribution, this excess is given by
\begin{eqnarray}
\Delta \Omega &=&\int d\vec{r} \frac{\delta\psi(\vec{r})}{2} \left(-\epsilon +3\bar{\psi}_l^2+(\nabla^2+1)^2\right)\delta\psi(\vec{r})\nonumber\\
&=&\int \frac{d\vec{k}}{2(2\pi)^3} \gamma(k) \delta\tilde{\psi}(\vec{k})\delta\tilde{\psi}(-\vec{k}),\nonumber\\
\label{fink}
\end{eqnarray}
where we have defined $\gamma(k)\equiv -\epsilon+3\bar{\psi}^2_l+(1-k^2)^2$ and
the Fourier transforms  
\begin{eqnarray}
\delta\tilde{\psi}(\vec{k}) &=&\int d\vec{r}\delta \psi(\vec{r})e^{-i\vec{k}\cdot\vec{r}},\\
\delta{\psi}(\vec{r}) &=&\frac{1}{(2\pi)^3}\int d\vec{k}\delta \tilde{\psi}(\vec{k}) e^{i\vec{k}\cdot\vec{r}}.
\end{eqnarray}

Next, the ``renormalized'' excess grand potential is the equilibrium statistical average of 
$\Delta \Omega$ given by 
\begin{equation}
\langle \Delta \Omega \rangle_{eq}=\int \frac{d\vec{k}}{2(2\pi)^3} \gamma(k) \langle \delta\tilde{\psi}(\vec{k})\delta\tilde{\psi}(-\vec{k})\rangle_{eq}.
\label{eqdomegaeq}
\end{equation}
The correlation function $\langle \delta\tilde{\psi}(\vec{k})\delta\tilde{\psi}(-\vec{k})\rangle_{eq}$ is readily obtained by
linearizing Eq. (\ref{fulldyn}) around the liquid state with the substitution $\psi(\vec{r},t) =\bar{\psi}_l+\delta\psi(\vec{r},t)$ 
and Fourier transforming the resulting equation, which yields
\begin{equation}
\partial_t\delta\tilde{\psi}(\vec{k}) = -\gamma(k)\delta\tilde{\psi}(\vec{k})+\tilde{\eta}(\vec{k},t).\label{linearized}
\end{equation}
Eq. (\ref{linearized}) has the solution
\begin{equation}
\delta\tilde{\psi}(\vec{k},t) = \int_0^t d\tau e^{-\gamma(k)(t-\tau)}\tilde{\eta}(\vec{k},\tau).
\label{eqtildepsik}
\end{equation}
Substituting this form into Eq. (\ref{eqdomegaeq}) and using the Fourier transform of Eq. (\ref{noisedef})
 \begin{equation}
\langle \tilde{\eta}(\vec{k},t) \tilde{\eta}'(\vec{k}',t') \rangle=\Gamma (2\pi)^{3}\delta(\vec{k}+\vec{k}')\delta(t-t'),
\end{equation}
we obtain
\begin{equation}
\langle\delta\tilde{\psi}(\vec{k},t)\delta\tilde{\psi}(\vec{k}',t)\rangle=\frac{(2\pi)^3\Gamma}{2\gamma(k)}\left(1-e^{-2\gamma(k)t}\right)\delta(\vec{k}+\vec{k}'),
\end{equation}
and thus in the equilibrium long time limit ($t \to \infty$)  
\begin{eqnarray}
\langle\delta\tilde{\psi}(\vec{k})\delta\tilde{\psi}(\vec{k}')\rangle_{eq} =&&\frac{(2\pi)^3\Gamma}{2\gamma(k)}\delta(\vec{k}+\vec{k}').
\end{eqnarray}
Substituting the above result into Eq.~(\ref{eqdomegaeq}), we obtain the final form of the excess liquid grand potential due to noise-induced density fluctuations
\begin{equation}
\langle \Delta \Omega \rangle_{eq}= \frac{4\pi \Gamma V}{4(2\pi)^3}\int_0^{k_{max}}k^2 dk 
= \frac{4\pi \Gamma V}{12(2\pi)^3}k_{max}^3.
\label{divergence}
\end{equation}
Eq.~(\ref{divergence}) shows that the renormalized grand potential of the liquid has an ultraviolet divergence that requires keeping $k_{max}$ finite. 
Next, to estimate what this cutoff should be,
we compare $\langle \Delta \Omega \rangle_{eq}$ to the barrier height of the double-well potential between solid and liquid. To compute the latter, we start from the expression
\begin{equation}
\Delta \Omega \equiv \Omega_s - \Omega_l,
\end{equation}
where
\begin{equation}
\Omega_l/V=  (-\epsilon+1)\frac{\bar{\psi}_l^2}{2}+\frac{\bar{\psi}_l^4}{4}-\mu\bar{\psi}_l,
\end{equation}
and $\Omega_s/V$ is obtained by substituting into Eq. (\ref{ffunc}) the form for a bcc crystal density field \cite{WuKarma2007} 
\begin{eqnarray}
& &\psi(\vec{r})-\bar{\psi}=\epsilon^{1/2} \sum_i A e^{i\vec{k}_i\cdot \vec{r}} = 4\epsilon^{1/2}A(\cos qx\cos qy\nonumber\\
& &+\cos qx\cos qz+\cos qy\cos qz),
\label{RLV}
\end{eqnarray}
where $\vec{k}_i$ are the set of 12 $[110]$ bcc principal reciprocal lattice vectors (cf, Eq. (\ref{rlvset}) below) and $q=1/\sqrt{2}$. This yields the expression
\begin{eqnarray}
& &\Omega_s/V = (-\epsilon+1)\frac{\bar{\psi}_s^2}{2}+\frac{\bar{\psi}_s^4}{4}-\mu\bar{\psi}_s\nonumber\\
& &-6\epsilon^2 A^2+18\epsilon\bar{\psi}_s^2A^2+48\epsilon^{3/2}\bar{\psi}_sA^3+135\epsilon^2A^4.
\end{eqnarray}
Expanding $\bar{\psi}_{s(l)}$ in powers of $\epsilon^{1/2}$ as $\bar{\psi}_{s(l)}=\psi_{s(l)0}\epsilon^{1/2}+\psi_{s(l)1}\epsilon+\dots$ and using the results (see Section III A of Ref.~\onlinecite{WuKarma2007}) that $\psi_{s(l)0}\equiv \psi_c=-\sqrt{45/103}$ and $\psi_{s(l)1}=0$,
we obtain the expression for the grand potential difference between solid and liquid per unit volume
\begin{equation}
\Delta \Omega = \epsilon^2Vf_{dw}(A)+O(\epsilon^3),
\end{equation}
where
\begin{equation}
f_{dw}(A)=-6A^2+18\psi_c^2A^2+48\psi_cA^3+135A^4\label{2well}
\end{equation}
has the shape of a double well potential with minima at $A_l=0$ and $A_s=8/(3\sqrt{515})$ corresponding to the liquid and solid states, respectively, and with a maximum at $A_b=4/(3\sqrt{515})$ corresponding to the free-energy barrier between those two states. Next, equating the renormalized excess grand potential of the liquid and the height of the double-well potential, $\langle \Delta \Omega \rangle_{eq}=\epsilon^2Vf_{dw}(A_b)$, and using Eqs. (\ref{Gammadef}) and (\ref{divergence}) together with the expressions for $g$, $\lambda$, and $\epsilon$ given in Appendix C, we obtain
\begin{equation}
k_{max}^3=\frac{3(2\pi)^2B\sqrt{-S(q_0)C''(q_0)^3}u_s^2n_0}{16\sqrt{2}}~f_{dw}(A_b), 
\end{equation}
where we have defined the dimensionless constant
\begin{equation}
B\equiv \frac{\sqrt{3\psi_c^2-1}}{\left(-\frac{2}{15}\psi_c+\frac{1}{15}\sqrt{5-11\psi_c^2}\right)^2}.
\end{equation}
Finally, evaluating $f_{dw}(A_b)$ using Eq.~(\ref{2well}) and using the parameters given in Table \ref{table3}, we find that $k_{max}\approx 0.918$, which implies that the shortest wavelength of the noise converted back to dimensional units is $\lambda_{min}=2\pi/(q_0k_{max})=0.77a$ where $a$ is the unit cell size.
Cutting off the noise at this lengthscale ensures that the noise does not wash out the barrier between the solid and liquid states. To be safe, we restrict ourselves to cutoffs on the order of one unit cell or larger;
details are presented in Appendix \ref{dynamicsA}. This choice clearly affects the relative impact of the noise, and we present results for two different choices of $\lambda_{min}$ below.

%%%%%%%%%%%%%%%%%%%%%%%%%%%%%%%%%%%%%%%
\section{Amplitude equations}
\label{gl::sec}

As a complementary approach we use here an amplitude equations (AE) model which is derived via a multiscale expansion from the three-dimensional phase-field-crystal model, Eq.~(\ref{ffunc}), and use $\epsilon$ as a small parameter in the regime of the phase diagram which describes the coexistence between the bcc and the homogeneous (melt) phase.

As in classical density-functional theory (DFT), the spatial variations of the density field, $\delta \psi(\vec{r})$, are expanded as a sum of density waves
\begin{equation}
\delta\psi(\vec{r})=\sum_j \uj e^{i\vec{k}^{(j)}\cdot\vec{r}},
\label{cexp}
\end{equation}
where $\vec{k}_i$ represent different reciprocal lattice vectors (RLVs) and $\uj$ are their associated amplitudes.
In the liquid phase, where the time averaged density is spatially constant, the amplitudes vanish, and in an undistorted solid phase they all attain the same constant value, $\uj=u_s$.

Unlike DFT, where a large number of modes is required to obtain sharp peaks around atomic positions, the simpler free-energy allows for a truncation of this sum to a small set of reciprocal lattice vectors.  Various methods have been developed to change the kernel of the free-energy in order to stabilize a variety of two and three dimensional periodic and crystal structures.
Here we focus on bcc structures, therefore we restrict the summation to the $N=12$ principal reciprocal vectors
\begin{eqnarray}
&&[110], [101], [011], [1\bar{1}0], [10\bar{1}], [01\bar{1}] \nonumber \\
&&[\bar{1}\bar{1}0], [\bar{1}0\bar{1}], [0\bar{1}\bar{1}], [\bar{1}10], [\bar{1}01], [0\bar{1}1]. \label{rlvset}
\end{eqnarray}
Notice that by the condition of having a real density field $\psi$ the $N$ complex amplitudes are not independent but are complex conjugate (denoted by a star) for antiparallel RLVs.
Therefore, we restrict the description to the first row of these RLVs and use only $N/2$ independent complex fields $\uj$.

A detailed derivation of the amplitude equations, which describe the evolution of the fields $u^{(j)}$ has been given in Refs.~\onlinecite{Wuetal2006,SpatschekKarma2010}, and therefore we only give the resulting expressions here.
The evolution equations can be derived from a free-energy functional, which reads
\begin{eqnarray}
F_{AE} &=& F_0 \int d\Rv \Bigg[ \sum_{i=1}^{N/2} |\Box_j \Aj |^2 + \frac{1}{12} \sum_{j=1}^{N/2} \Aj\Ajs \nonumber \\
&& + \frac{1}{90} \Bigg\{ \left( \sum_{j=1}^{N/2} \Aj\Ajs \right)^2  - \frac{1}{2}\sum_{j=1}^{N/2} |\Aj|^4 \nonumber \\
&& + 2A_{110}^* A_{1\bar{1}0}^* A_{101} A_{10\bar{1}} + 2A_{110} A_{1\bar{1}0} A_{101}^* A_{10\bar{1}}^* \nonumber \\
&& + 2A_{1\bar{1}0} A_{011} A_{01\bar{1}} A_{110}^* + 2A_{1\bar{1}0}^* A_{011}^* A_{01\bar{1}}^* A_{110} \nonumber \\
&& + 2A_{01\bar{1}} A_{10\bar{1}}^* A_{101} A_{011}^* + 2A_{01\bar{1}}^* A_{10\bar{1}} A_{101}^* A_{011} \Bigg\} \nonumber \\
&& -\frac{1}{8} \big( A_{011}^* A_{101} A_{1\bar{1}0}^* + A_{011} A_{101}^* A_{1\bar{1}0} + A_{011}^* A_{110} A_{10\bar{1}}^* \nonumber \\
&& + A_{011} A_{110}^* A_{10\bar{1}} + A_{01\bar{1}}^* A_{110} A_{101}^* + A_{01\bar{1}} A_{110}^* A_{101} \nonumber \\
&& + A_{01\bar{1}}^* A_{10\bar{1}} A_{1\bar{1}0}^* + A_{01\bar{1}} A_{10\bar{1}}^* A_{1\bar{1}0} \big) \Bigg] + F_T. \label{fenAE}
\end{eqnarray}
Here, we have introduced rescaled amplitudes
\begin{equation}
\Aj = \uj/u_s
\end{equation}
and a dimensionless ``slow'' scale $\Rv$ which is introduced below.

In a more general context the above expression can be derived from classical DFT and yields
\begin{eqnarray} 
F_{AE} &=& \frac{n_0 k_B T}{2} \int d\rv \Big( \sum_j \left[ \frac{\uj \ujs}{S(q_0)}- \frac{C''(q_0)}{2} \left|\square_j \uj\right|^2 \right] \nonumber \\ 
&& + f(\{\uj\}, T) \Big),
\end{eqnarray}
where the function $f(\{\uj\}, T)$ contains the higher order nonlinear terms in the amplitudes $\uj$ and an explicit dependence on the temperature $T$.
The form of these nonlinearities depends slightly on the underlying model:
Above it is given for a PFC model, and there are some differences if these terms are derived from DFT using an equal weight ansatz.
However, the differences are small and lead e.g. only to tiny changes of the anisotropy of the surface energy, as had been investigated in \cite{WuKarma2007}.
Furthermore, $n_0$ is the density in the liquid state, $C(r)$ is the direct correlation function with Fourier transform
\begin{equation}
C(q) = n_0 \int d\rv\, C(r) \exp(-i\kv\cdot\rv)
\end{equation}
with $r=|\rv|$ and $q=|\kv|$;
it is related to the liquid structure factor by
\begin{equation}
S(q) = \frac{1}{1-C(q)}.
\end{equation}
Here, all quantities are evaluated at the (first) peak of the structure factor $q_0$.
We introduce a (dimensionless) length scale
\begin{equation}
\Rv =\epsnew^{1/2}q_0 \rv
\end{equation}
with
\begin{equation} \label{gl::eq1}
\epsnew = - \frac{24}{S(q_0) C''(q_0) q_0^2} = \frac{96}{103} \epsilon.
\end{equation}
This allows the identity
\begin{equation}
F_0 = - \frac{n_0 k_B T}{2} C''(q_0) q_0^{-1} u_s^2 \epsnew^{-1/2}.
\end{equation}
The differential operator $\Box_j$ is given by
\begin{equation}
\Box_j = \khj\cdot \nabla - \frac{i\epsnew^{1/2}}{2} \nabla^2,
\end{equation}
where the nabla operator acts on the slow scale $\Rv$, and the vectors $\khj$ are the normalized principal RLVs.
The second term in the operator preserves the rotational invariance of the equations and is related to the use of the nonlinear strain tensor in elasticity.

The thermal tilt 
\begin{equation} \label{tilt::eq4}
F_T = L \frac{T-T_M}{T_M} \epsnew^{-3/2} q_0^{-3} \int d\Rv  \sum_{j=1}^{N/2} \frac{2\sqrt{\uj \ujs}}{N u_s}
\end{equation}
is added phenomenologically to the model.
Here, $T_M$ is the melting temperature, $L$ the latent heat and $\phi$ is an ``order parameter'' which discriminates between solid and liquid, with
\begin{equation} \label{tilt::eq3}
\phi(\{\Aj\}) = \frac{2}{N} \sum_{j=1}^{N/2}     \sqrt{\Aj \Ajs}.
\end{equation}
We note that this expression is invariant under elastic deformations and lattice rotations, which affect the complex phases of the amplitudes.
The choice of the coupling function and its implications are discussed in more detail in Ref.~\onlinecite{ShiftedCrystalPaper}.

Thermodynamic equilibrium corresponds to a stationary state of the free-energy functional, and we use relaxation dynamics according to
\begin{equation}
\frac{\partial \Aj}{\partial t} = - K_j \frac{\delta F}{\delta \Ajs}.
\end{equation}
Since we focus here exclusively on static situations, the choice of the kinetic coefficients $K_j$ is arbitrary.

This description predicts the correct anisotropy of surfaces energies \cite{Wuetal2006} and elastic properties and contains naturally the linear theory of elasticity.
The above Ginzburg-Landau model for crystals is conceptually close to theories of superconductivity and pattern formation in hydrodynamics \cite{CrossHohenberg1993}.

%%%%%%%%%%%%%%%%%%%%%%%%%%%%%%%%%%%%%%%%%%%%%%%%%%%%%%%%%%%
\section{Grain boundary premelting: Methodology}
\label{methodology::sec}

%%%%%%%%%%%%%%%%%%%%%%%%%%%%%%%%%%%%%%%%%%%%%%%%%%%%%%%%%%%
\subsection{Disjoining Potentials without noise}

While the definition of the disjoining potential in the introduction of this paper is framed in terms of a Gibbs ensemble (constant T, p, and N), it is easier in the PFC model to conduct simulations in the grand canonical ensemble (constant T, $\mu$, and V). As is discussed in Ref.~\onlinecite{Mellenthinetal2008}, studying grain boundary behavior as a function of $\mu-\mu_{eq}$ is equivalent to using $T-T_M$.
In the spirit of Eq.~(\ref{gbexc}) we can define the disjoining potential as the excess grand potential energy in the system
\begin{equation}
\Omega(\mu)=L_yL_z[(L_x-W(\mu))\omega_s(\mu)+W(\mu)\omega_l(\mu)+2\gamma_{sl}+V(W)],
\label{totalgb}
\end{equation}
where $\omega_{s(l)}(\mu)$ are the grand potential densities of of the solid (liquid) respectively, $W$ is the width of the liquid layer, $L_yL_z$ is the area of the grain boundary and $L_x$ is the length of the system perpendicular to the boundary.

We can then define the grain boundary energy
\begin{eqnarray}
\gamma_{gb}(\mu)=&&\frac{\Omega(\mu)-L_xL_yL_z\omega_s(\mu)}{L_yL_z}=(\omega_l(\mu)-\omega_s(\mu))W(\mu)\nonumber\\
&&+2\gamma_{sl}+V[W(\mu)],\label{ggb}
\end{eqnarray}
In order to calculate the liquid layer width $W$ we relate it to the excess mass in the system due to the presence of a grain boundary. We define an excess mass per unit area of the grain boundary by comparing the mass of two systems at the same chemical potential, one a perfect solid and one containing a grain boundary,
\begin{equation}
\psi_{exc}(\mu)=L_x[\bar{\psi}(\mu)-\bar{\psi}_s(\mu)].
\end{equation}
By also obtaining the liquid density at the same chemical potential we can say that the excess mass must be equal to the liquid thickness times the difference in densities of the liquid and solid phases 
\begin{equation}
W[\bar{\psi}_l(\mu)-\bar{\psi}_s(\mu)]=\psi_{exc}(\mu).
\end{equation}
Combining the two equations gives a definition for the liquid layer $W(\mu)$,
\begin{equation}
W(\mu) = L_x\frac{\bar{\psi}(\mu)-\bar{\psi}_s(\mu)}{\bar{\psi}_l(\mu)-\bar{\psi}_s(\mu)}.
\end{equation}
In order to calculate the disjoining potential from Eq.~(\ref{ggb}), we need a way to calculate the grain boundary energy $\gamma_{gb}$. By dividing Eq.~(\ref{totalgb}) by $L_xL_yL_z$ and using Eq.~(\ref{ggb}) we see that

\begin{equation}
\omega=\omega_s(\mu)+\gamma_{gb}(\mu)/L_x.
\label{gbextract}
\end{equation}

\begin{figure}
\centering
\includegraphics[width=0.5\textwidth, angle=0]{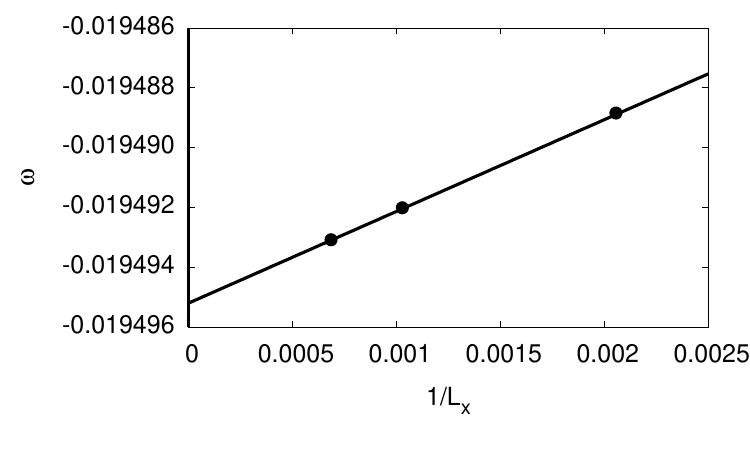}
\caption{Grand potential energy density $\omega$ at a constant chemical potential $\mu=-.19609$ for three different inverse system sizes $1/L_x$ perpendicular to the grain boundary for a misorientation of $\theta=18.9^{\circ}$.  The intercept is twice the grain boundary energy due to periodicity.}
\label{inv}
\end{figure}
As seen in Fig. \ref{inv} we can then obtain the grain boundary energy by running a series of simulations at fixed chemical potential and varying system size. Plotting $\omega$ as a function of the inverse system size, we see a clear linear relationship, and are able to extract $\gamma_{gb}$.  We should also note that as the system is periodic we have two identical grain boundaries and the resulting energy should be divided by two. Theoretically, all that is needed to compute the disjoining potential is to calculate the grain boundary energy for a whole range of chemical potentials, and then use Eq.~(\ref{ggb}) to extract the disjoining potential for each $\mu$. However, in practice this is computationally unfeasible. Instead we choose an alternate route and define the disjoining pressure $\Pi$ 
\begin{equation}
\Pi=-\frac{1}{L_yL_z}\frac{\partial\Omega}{\partial W}=\omega_s-\omega_l-V'(W).
\label{pressure}
\end{equation}
We also recognize that the grand potential densities $\omega$ can be expanded as a function of $\mu$,
\begin{equation}
\omega_s-\omega_l\approx-(\bar{\psi}_s^{eq}-\bar{\psi}_l^{eq})(\mu-\mu_{eq}).
\end{equation}
When the system is at equilibrium $\Pi$ is equal to zero and
\begin{equation}
V'(W(\mu)) = -(\bar{\psi}_s^{eq}-\bar{\psi}_l^{eq})(\mu-\mu_{eq}).
\label{vinteg}
\end{equation}
As noted in Ref.~\onlinecite{Mellenthinetal2008}, $V'(\mu)$ is a known function of $\mu$ which depends only on bulk thermodynamics. By conducting simulations for a range of fixed chemical potentials we can obtain $V'(W)$ which is then integrated to get $V(W)$ as long as we know $\gamma_{gb}$ for one particular $\mu$ (calculated from Eq.~(\ref{gbextract})). This method is far more computationally tractable than calculating $V(W)$ directly from Eq.~(\ref{ggb}).

While this discussion as focused on undercooling as a function of chemical potential $\mu$, it is possible to relate this quantity to undercooling as a function of temperature. As derived in the supplementary material of Ref.~\onlinecite{Olmstedetal2011} we relate the energy difference between solid and liquid phases to the difference in grand potential energy densities and expand $\omega$ in $\mu$
\begin{equation}
L\frac{T-T_M}{T_M}=\omega_s-\omega_l\approx-(\bar{\psi}_s^{eq}-\bar{\psi}_l^{eq})(\mu-\mu_{eq}),
\label{Tconv}
\end{equation}
where the latent heat of fusion for the  MD Fe potential is $L=1.968\cdot10^{-24}\,\textrm{kJ}/\textrm{\AA}^3$.  This allows us to present PFC results as a function of temperature rather than chemical potential.
%
%\begin{eqnarray}
%L&=&15.63[\textrm{kJ}\cdot \textrm{mol}^{-1}]\\
%&=&1.99\cdot10^{-24}[\textrm{kJ}/\textrm{\AA}^3].
%\end{eqnarray}
%We can now relate the undercooling $\mu$ to a real temperature by using,
%
%\begin{eqnarray}
%&&\frac{T-T_m}{T}1.99\cdot 10^{-24}[\textrm{kJ}/\textrm{\AA}^3]\nonumber\\
%&&=-(\bar{\psi}_s^{eq}-\bar{\psi}_l^{eq})(\mu-\mu_{eq})8.81\cdot10^{-21}[\textrm{kJ}/\textrm{\AA}^3],
%\end{eqnarray}
%\begin{equation}
%\frac{T-T_m}{T}=-(\bar{\psi}_s^{eq}-\bar{\psi}_l^{eq})(\mu-\mu_{eq})\cdot4427.5.
%\end{equation}
%
%

For the amplitude equations we use a similar approach to define the melt layer thickness and the disjoining potential.
Here we focus on the case that the thermal tilt in the free energy is a linear function in amplitudes, which has the consequence that the bulk amplitudes in solid and liquid deviate from 0 and 1, respectively, which are the potential minima for $T=T_M$.
In the bulk all amplitudes have the same magnitude $A$ (they differ only in phase due to the grain rotation), and therefore the free energy density becomes
\begin{eqnarray*}
f_{AE}(A, T) &=& f_0(A) + f_T \\
&=& F_0 \epsnew^{3/2} q_0^3 \left( \frac{1}{2} A^2 - A^3 + \frac{1}{2} A^4 \right) + A L \frac{T-T_M}{T_M}.
\end{eqnarray*}
Notice that in the bulk the gradient square term does not appear despite the spatial variations due of the complex phases;
this is a property of the box operator, as discussed in detail in Ref.~\onlinecite{SpatschekKarma2010}.
The minimum positions of the bulk free energy density can be found as the roots of the cubic equation $f_0'(A) + f_T'(A) =0$, and we denote them by $A_l(T)$ and $A_s(T)$ (the third root is the maximum in between these two minima).
We define the melt layer thickness for a two-dimensional system as
\begin{equation}
W = q_0^{-1} \tilde{\epsilon}^{-1/2} \frac{1}{N L_y L_z} \sum_{j=1}^N \int d\Rv \frac{|\Aj(\Rv)| - A_s}{A_l-A_s},
\end{equation}
where the interface area $L_yL_z$ is measured on the slow dimensionless scale.
The prefactor $q_0^{-1}\tilde{\epsilon}^{-1/2}$ stems from the conversion of $W$ to the dimensional ``short'' scale.
We let the simulations relax towards equilibrium with the bias potential $F_T$, which does not enter into the expression of the disjoining potential $V(W)$.
The latter is calculated as
\begin{eqnarray}
V(W) &=& \frac{F_0 q_0^2\tilde{\epsilon}}{L_yL_z} \Bigg[ \int d\Rv f  - L_yL_z W f(A_l) + \nonumber \\
&& L_yL_z (L_x - W) f(A_s) \Bigg] - 2 \gamma_{sl},
\end{eqnarray}
in analogy to Eq. (\ref{totalgb}) above.

By tuning the bias potential $F_T$ we can therefore extract the convex parts of the disjoining potential, as before for the PFC simulations;
this approach is only possible in the convex regions, $V''(W)>0$, otherwise the equilibrium state is unstable.
To also get estimates for the concave parts for the attractive potentials we instead do dynamical runs at $T=T_M$, starting with a rather wide liquid layer.
Due to the attractive interaction the melt starts to solidify, and we measure $W$ and $F$ as function of time.
Although the system is not in full equilibrium during this evolution, the extracted disjoining potential $V(W)$ matches well to the static results which were obtained by the above method.

%%%%%%%%%%%%%%%%%%%%%%%%%%%%%%%%%%%%%%%%%%%%%%%%%%%%%%%%%%%
\subsection{Thermal fluctuations}
\label{thermfluc}
Calculating the disjoining potential from a fluctuating interface requires a different technique then presented above for PFC without noise.  Instead we follow the approach of Hoyt et al.~\cite{Hoytetal2009, Fensinetal2010}, which relates the excess interfacial free-energy to fluctuations of the liquid layer width as a function of undercooling. The grain boundary free-energy $\gamma_{gb}(\mu)$ is defined as above in Eq.~(\ref{d9n}), allowing us to easily write the probability of observing a liquid layer width $W$ as 
\begin{equation} \label{fluc::eq1}
P(W,T) = C \exp[-A\gamma_{gb}(T)/k_BT],
\end{equation}
where $A$ is the interface area, $C$ is a temperature dependent normalization constant, and we have used Eq.~(\ref{Tconv}) to convert from $\mu$ to $T$.  We obtain these probability distributions by conducting simulations with noise for various undercoolings and plotting the resulting histograms, as seen in Fig.~\ref{hk1}. The inclusion of noise prevents us from using the method presented above to calculate the liquid layer width, due to the large fluctuations in average density $\bar{\psi}$. Therefore we use the procedure presented below in Appendix \ref{wnoise} to define the width $W$. Once we have the width histograms we can extract the disjoining potential by writing
\begin{equation} \label{fluc::eq2}
V(W)=-(k_BT/A)\ln P(W,T_i)-\Delta \omega_f W-2\gamma_{sl}+a_i,
\end{equation}
where $i$ enumerates simulations run at different temperatures $T_i$, $\Delta \omega_f$ is the bulk grand potential energy density difference between solid and liquid for a particular undercooling, and the $a_i$ are free constants. The $a_i$ are then used to match the various disjoining potential curves (from different undercoolings) into one continuous disjoining potential as seen in Fig.~\ref{d9n}.

%%%%%%%%%%%%%%%%%%%%%%%%%%%%%%%%%%%%%%%%%%%%%%%%%%%%%%%%%%%
\subsection{Molecular Dynamics simulations}

For the purpose of comparing with PFC and AE predictions, MD calculations of disjoining potentials have been undertaken for bcc Fe using the method described in Refs.~\onlinecite{Hoytetal2009, Fensinetal2010}.
This approach for computing $V(W)$ by MD involves the use of equilibrium MD simulations to compute the grain-boundary width histograms, from which the disjoining potential can be extracted using equations analogous to Eqs.~(\ref{fluc::eq1}) and (\ref{fluc::eq2}) (see Ref.~\onlinecite{Hoytetal2009} for details). 
In the current application of this approach we performed MD simulations using the LAMMPS code \cite{Plimpton95}. 
The simulations were performed in the $NP_zAT$ ensemble, corresponding to fixed particle number $N$ and cross-sectional area $A$, with the temperature $T$ and pressure normal to the grain boundary $P_z=0$ controlled by a Nos\'{e}-Hoover thermostat and barostat \cite{NoseHoover}. 
Use was made of periodic boundary conditions, such that the simulation system contained two identical symmetry tilt boundaries. 
Interatomic interactions were modeled by the embedded-atom-method potential given in Ref.~\onlinecite{Mendelevetal2003}. 
Width histograms were computed from snapshots of equilibrium MD simulations lasting at least 80 ns, at two or more temperatures below, but within 10 K, of the melting point. 
For each snapshot, the grain-boundary width was determined through the use of a structural order parameter based on a formulation due to Morris \cite{Morris}. 
Further details can be found in the supplementary information to Ref.~\onlinecite{Olmstedetal2011}.

%%%%%%%%%%%%%%%%%%%%%%%%%%%%%%%%%%%%%%%%%%%%%%%%%%%%%%%%%%%
\section{Grain Boundary Premelting: Results}
\label{gbpremelting::sec}

In this section we discuss the results for grain boundary premelting in the context of phase field crystal (PFC) and amplitude equations (AE) modeling.
In particular, we inspect symmetric tilt grain boundaries in bcc $\delta$ iron for misoriented [100] surfaces.
The results are compared to MD simulations.

There are two relevant sets of information that are extracted here. 

First, the width of the melt layer $W$ forming at the grain boundary as function of temperature is discussed.
In thermodynamic equilibrium, an infinitely wide liquid phase can coexist with the two grains, and in this limit interactions between the grains with different orientation have decayed completely.
Therefore, the curves $W(T)$ must diverge at $T=T_M$.
For a repulsive grain-grain interaction this leads to monotonously increasing and diverging functions $W(T)$ from below the melting point.
For attractive grain boundaries, the dependence is not unique, since the diverging branch is located above the melting point and unstable.
Additionally, a stable branch with finite melt layer thickness at $T=T_M$ exists with lower $W$, and it merges with the unstable branch at a finite overheating temperature $T_o>T_M$.
Above the melting point this branch is only metastable, as from bulk thermodynamics the system would reduce its total free energy by complete melting.
Here, we only track the (meta-)stable branches of the solution, as they can be obtained by simple gradient dynamics.
For a more extended discussion and the consideration of unstable branches in the framework of phase field modeling we refer to Ref.~\onlinecite{Wangetal2008}.
In contrast to this kind of modeling, which is characterized by a constant order parameter in each bulk phase, we preserve in the PFC and AE simulations the atomic structure, and therefore the interaction can be studied as function of the grain misorientation, and one finds a transition from attractive situations for low angle grain boundaries to repulsive interactions for high angle boundaries.

Second, related to the $W(T)$ curves is the disjoining potential $V(W)$, which describes the interaction between the grains which are separated by the melt layer of thickness $W$.
The disjoining potential is normalized such that it decays to zero for $W\to\infty$, when the excess energy at the melting point reduces to twice the solid-liquid interfacial energy per unit area, $\gamma_{sl}$.
A repulsive interaction corresponds to $V'(W)<0$, and an attractive case to $V'(W)>0$.
Typically, an attractive situation at large distances still comes with a repulsive interaction if the interfaces are very close to each other.

In subsection \ref{deterministic} we present the deterministic results of the continuum simulations, i.e.~without thermal fluctuations.
The deviations from the MD results, which show a transition from attraction to repulsion at lower angles, is interpreted as a different ratio  of the solid-liquid interfacial energy $\gamma_{sl}$ and the grain boundary energy $\gamma_{gb}$.
Therefore, in subsection \ref{epsrescaling}  this ratio is set to a similar value in the continuum theories, leading to a shift of the curves in the right direction.

Finally, in subsection \ref{fluctuations} the influence of thermal noise is studied in the PFC model, which also leads to a shift of the disjoining potential closer to the atomistic results.

%%%%%%%%%%%%%%%%%%%%%%%%%%%%%%%%%%%%%%%%%%%%%%%%%%%%%%%%%%%
\subsection{AE, PFC without noise and MD results}
\label{deterministic}

\begin{figure}
\centering
\includegraphics[width=0.5\textwidth, angle=0]{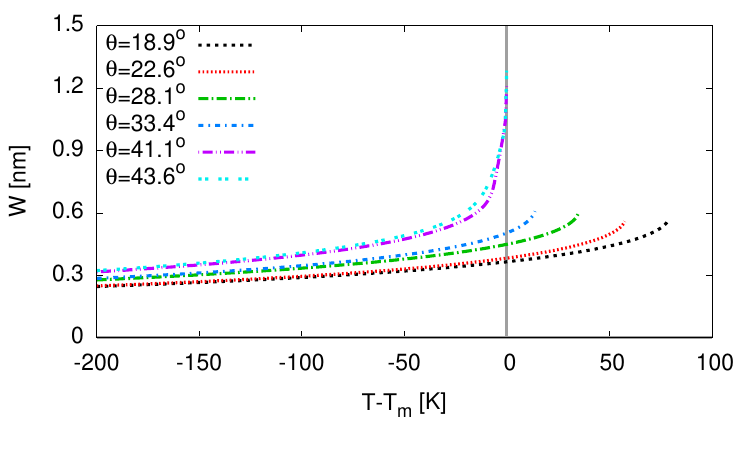}
\caption{(Color online) Grain boundary liquid layer width as a function of undercooling for various misorientations at a value of $\epsilon=0.0923$.  There is a transition from attractive to repulsive grain boundaries at approximately $\theta=37^{\circ}$. Data is from PFC simulations.}
\label{w9}
\end{figure}

The PFC results for liquid layer width as a function of undercooling are presented in Fig \ref{w9}.  As detailed above, for each misorientation the system is relaxed at constant chemical potential $\mu$ and the liquid layer width is measured.  Then $\mu$ is increased and the system is relaxed again, until the amount of liquid diverges.  PFC simulations are conducted with $\Delta t = 0.5$, and $\Delta x \approx a/8$ where $a$ is one lattice spacing.  In order to ensure periodic simulation geometry for rotated bi-crystals, we follow the procedure set forth in Appendix B of Ref.~\onlinecite{Mellenthinetal2008}.  The system size in the direction normal to the boundary ($L_x$) is kept as close to $L_x\approx100a$ as possible given the constraints of periodicity, while the system size along the boundary is set to the minimum length allowed under periodicity and $L_z=2a$.

There are qualitative similarities and differences to sharp interface theory.  
Here we find that, for the repulsive boundaries, there is indeed a logarithmic divergence of the melt layer thickness if the temperature reaches $T_M$ from below, which indicates short range interactions.  For low temperatures, there is no bridging temperature.  Instead there is always a small liquid layer in existence.  
%For the attractive boundaries, the maximum overheating $T_o$ temperature is  greater than the melting temperature, and the system is metastable for $T_M<T<T_o$.

\begin{figure}
\centering
\includegraphics[width=0.5\textwidth, angle=0]{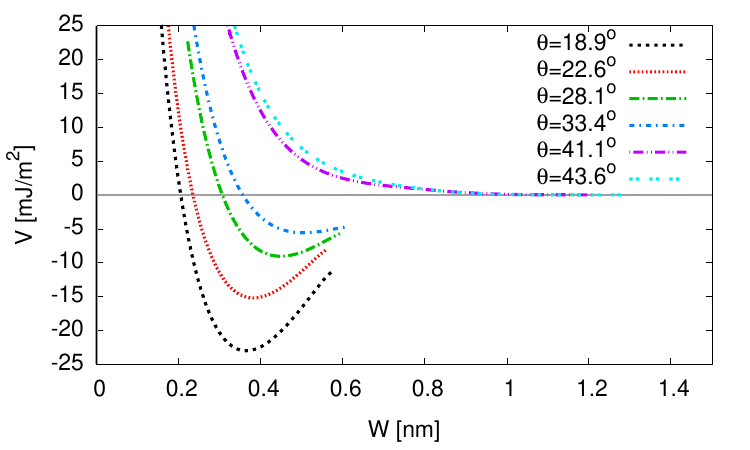}
\caption{(Color online) Disjoining potential as a function of grain boundary width for various misorientations and $\epsilon=0.0923$ from PFC simulations.  There is a transition from attractive to repulsive disjoining potentials at approximately $\theta=37^{\circ}$.}
\label{disj9}
\end{figure}

The disjoining potentials as integrated from Eq.~(\ref{vinteg}) for $\epsilon=0.0923$ are plotted in  Fig \ref{disj9}.  Here we can easily see the transition between repulsive and attractive disjoining potentials at $\theta\approx37^{\circ}$.
% Data is from PFC simulations..

\begin{figure}
\begin{center}
\includegraphics[width=9cm]{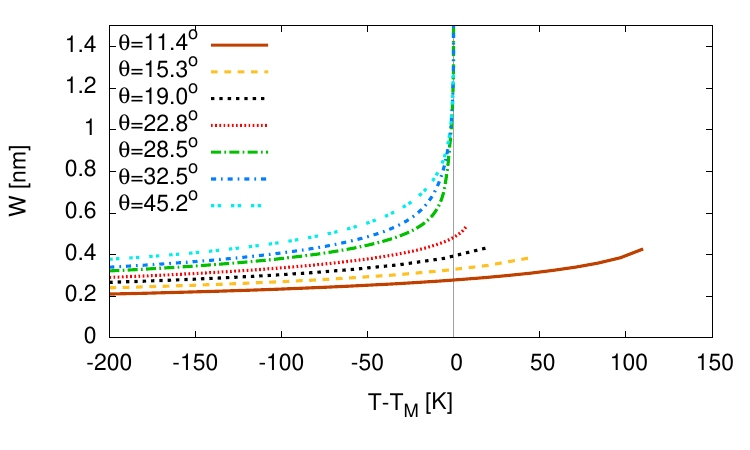}
\caption{(Color online) Width of the melt layer as function of the deviation from the melting temperature for $\epsilon=0.0923$, as obtained from AE.
}
\label{gb::fig3}
\end{center}
\end{figure}

\begin{figure}
\begin{center}
\includegraphics[width=9cm]{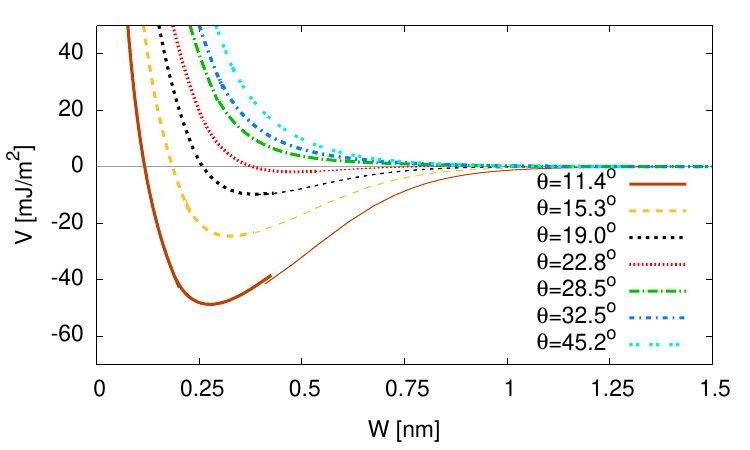}
\caption{(Color online) Disjoining potential for inclined (100) interfaces for $\epsilon=0.0923$, obtained from AE. Below a critical misorientation $\theta_c\approx 25^\circ$ the interfaces attract each other, above this value they are repulsive. 
The thin (concave) parts of the attractive potentials are obtained by dynamical simulations.
}
\label{gb::fig4}
\end{center}
\end{figure}

\begin{figure}
\begin{center}
\includegraphics[width=9cm]{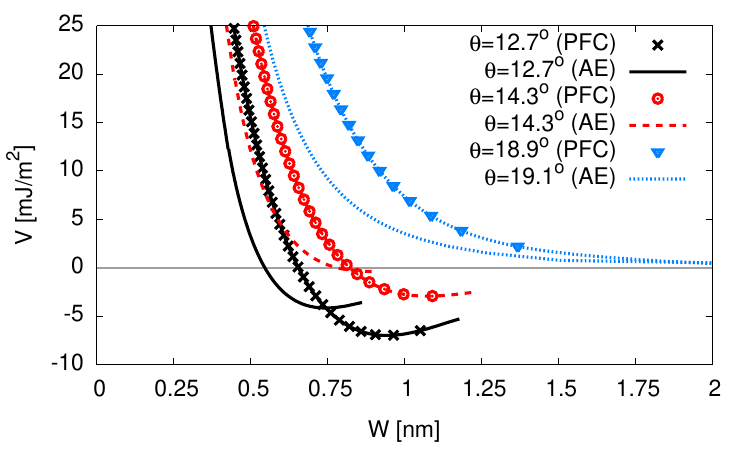}
\caption{(Color online) Disjoining potential comparison for $\epsilon=0.0923$ for selected misorientations for PFC and AE data.}
\label{gl::fig5}
\end{center}
\end{figure}

The corresponding results for AE are shown in Figs.~\ref{gb::fig3} and \ref{gb::fig4}.
Here we typically use a lattice spacing of $\Delta x \approx 0.5$ (measured on the dimensionless slow scale) in a spectral code.
They show a qualitatively similar behavior, and in particular the results of this coarse-grained theory deliver results which are on the same order of magnitude.
The main difference is that the behavior is already significantly more repulsive, and the transition between attraction and repulsion occurs around $\theta=25^\circ$.
We also show here results of the concave parts of the disjoining potential curves, $V''(W)<0$, which were obtained by using dynamical simulations.
As mentioned before, dynamic and static simulations deliver comparable results in the range where both of them apply;
the small discrepancy can be seen best for the $11.4^\circ$ misorientation, where only a small mismatch of the order of $1\,\mathrm{mJ/m}^2$ exists.
Similarly to the stronger repulsion, the overheating range, i.e.~the temperatures up to which a bicrystal is still thermodynamically metastable, is lower accordingly.
A detailed comparison of the disjoining potentials, which were computed by these two continuum models is shown in Fig.~\ref{gl::fig5}.
It turns out that the positions of the energetic minima are very similar, although the definition of $W$ is not the same in the two models, and therefore horizontal shifts of the curves would be conceivable.
Nevertheless, the energy values, in particular in the minima, are still quite different for low angle misorientations.

We point out that this disagreement is not surprising, as AE is derived from the PFC model by a multiscale expansion, which assumes that the length scale over which the amplitudes vary spatially is large compared to the lattice unit.
This assumption breaks down for a high angle grain boundary, which introduces a short-scale oscillation of the amplitudes, and therefore discrepancies between the models have to be expected, as discussed in detail in Ref.~\onlinecite{SpatschekKarma2010}.
%A high angle rotation, however, introduces a most oscillating factor to the amplitudes, and therefore the discrepancies are expected.

\begin{figure}
\centering
\includegraphics[width=0.5\textwidth, angle=0]{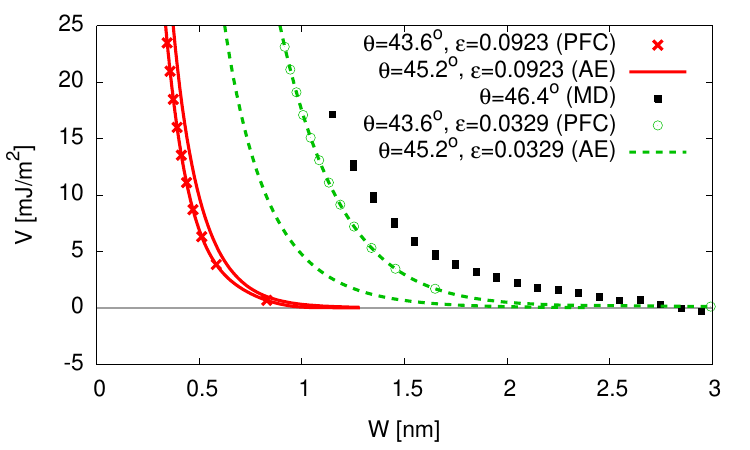}
\caption{(Color online) A comparison of disjoining potentials between PFC, AE and MD for a misorientation of  $\theta \approx44^{\circ}$, with $\epsilon=0.0923$ and $\epsilon=0.0329$. The MD results represent statistics generated from three different undercoolings.
}
\label{compare9}
\end{figure}

A comparison with the MD data in Fig.~\ref{compare9} shows that the two continuum theory curves are very close together for a high angle misorientation of about $45^\circ$, but still quantitatively different from  the atomistic results.
The reasons for this disagreement will be elucidated in more detail in the next section.
%Additionally, the definitions of $W$ for the different methods may partially explain the discrepancy between the results.
%We stress that again the definition of $W$ might be quite different between the methods, and lateral shifts of the curves may be appropriate.

%%%%%%%%%%%%%%%%%%%%%%%%%%%%%%%%%%%%%%%%%%%%%%%%%%%%%%%%%%%
\subsection{Rescaling of $\epsilon$}
\label{epsrescaling}

As seen in Fig. \ref{compare9}, although qualitatively similar, there is a large difference in the range of interaction between MD and PFC.  

\begin{figure}
\centering
\includegraphics[width=0.5\textwidth, angle=0]{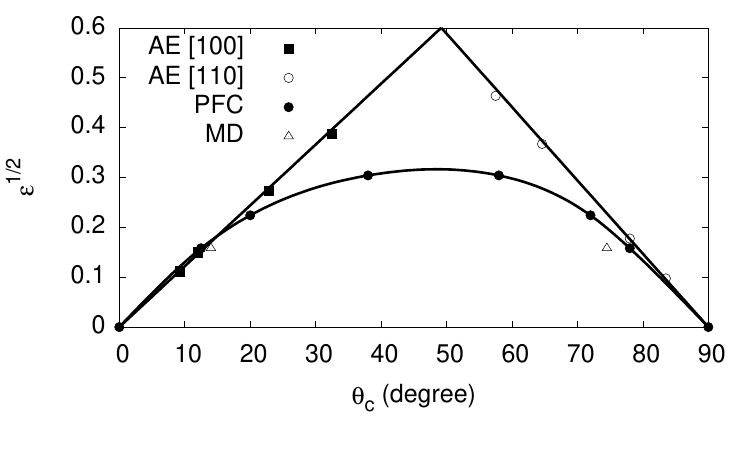}
\caption{Plot of critical angle $\theta_c$ as a function of $\epsilon^{1/2}$ for both PFC and AE.  Disjoining potentials are repulsive (attractive) below (above) those lines. The range of misorientation between triangles where GBs are observed to form a liquid-like layer in MD simulations \cite{Olmstedetal2011} is in good quantitative agreement with the range predicted by PFC and AE for the value of $\epsilon$ that matches the ratio of bulk modulus and solid-liquid interfacial free-energy in MD simulations. The AE theory predicts the asymptotic form $\theta_c\sim \epsilon^{1/2}$ (also $\pi-\theta_c\sim \epsilon^{1/2}$ for misorientations close to 90$^0$) for $\epsilon\ll 1$.}
\label{nose}
\end{figure}

Here we aim at a better matching of the continuum and the atomistic simulations.
To this end, we first investigate the critical angle for the transition between attraction and repulsion as function of the only free parameter, which appears in the (rescaled) PFC and AE models, namely $\epsilon$.
This dependence is shown in Fig.~\ref{nose}.
%
%In order to facilitate a better comparison between the two models, we must look to the only adjustable parameter, $\epsilon$.
%

The observation is that for low values of $\epsilon$ the critical angle scales as $\theta_c\sim \epsilon^{1/2}$ both for PFC and AE, and only for higher angles the results differ from each other.
In particular, there is no premelting transition in PFC for $\epsilon^{1/2} \gtrsim 0.3$, in agreement with findings in Ref.~\onlinecite{Berryetal2008}.
On the other hand, the critical misorientation curves are two separate straight lines, coming from (100) and (110) interfaces;
this is a consequence of the incorrect treatment of the (discrete) rotational invariance of the model, as discussed in detail in Ref.~\onlinecite{ShiftedCrystalPaper}.

The basic physical argument for the premelting transition is that $2\gamma_{sl}=\gamma_{gb}$.
The scaling of the surface energy is
\begin{equation} \label{gammascaling}
\gamma_{sl} \sim n_0 k_B T \left(\frac{-C''(q_0)}{S(q_0)} \right)^{1/2} u_s^2.
\end{equation}
The grain boundary energy, on the other hand, stems from the elastic energy of the geometrically necessary dislocations at the grain boundary.
For fixed misorientation the strain is therefore fixed, and the elastic energy scales with the elastic constants.
Thus we obtain
\begin{equation} 
B \sim n_0 k_B T (-C''(q_0)) q_0^2 u_s^2,
\end{equation}
with $B$ being the bulk modulus.
The ratio of these two parameters therefore scales as
\begin{equation} \label{ratioGgamma}
\frac{B}{\gamma_{sl}} \sim q_0\epsilon^{-1/2}.
\end{equation}
For low misorientations the grain boundary energy depends on $\theta$ as given by the Read-Shockley law,
\begin{equation}
\gamma_{gb} = \gamma_0 \theta (A - \ln\theta) \sim \gamma_0 \theta, 
\end{equation}
where $\gamma_0$ and $A$ depend on the elastic constants, the Burgers vector and the dislocation core radius;
the second relation is the asymptotic behavior for small misorientations.
Comparing this to Eq.~(\ref{ratioGgamma}) therefore implies for the critical misorientation with $\gamma_{gb}\sim 2\gamma_{sl}$
\begin{equation}
\theta_c \sim \epsilon^{1/2}\; \mbox{ for } \epsilon\ll 1.
\end{equation}

A more geometrical interpretation of the transition between attraction and repulsion stems from a length scale comparison.
For low angle grain boundaries, the spacing between neighboring dislocations scales as $1/q_0\theta$.
On the other hand, a dislocation is a defect, which corresponds to a singularity in the continuum theories.
For a Burger's circuit around the dislocation line the phase of the complex amplitudes increases by a multiple of $2\pi$, and at the same time density and the magnitude of the amplitudes tends to zero inside the dislocation (or at least they become small).
Therefore, a dislocation core is similar to a solid-liquid interface, and its diameter is therefore proportional to the extent of a solid-liquid interface, i.e.~$q_0^{-1}\epsilon^{-1/2}$.
The transition between attraction and repulsion happens when the dislocation spacing becomes comparable to the dislocation core size, thus $\theta_c\sim \epsilon^{1/2}$.

In AE, which is derived from the PFC model in the limit of small values of $\epsilon$, the scaling $\theta_c\sim \epsilon^{1/2}$ can also be inspected from another perspective.
For small $\epsilon$ the correction term in the box operator becomes negligible, and therefore $\epsilon$ drops out of the dimensionless bulk equations.
It only enters into the problem by the geometrical setup of the grain boundary, which requires the rotation of the crystals.
This leads to amplitudes 
\begin{equation}
A_j = \exp[i \vec{k}_j^\dagger ({\mathbf{R} - \mathbf{1}) \vec{R}/\tilde{\epsilon}^{1/2}}]
\end{equation}
on the slow scale $\vec{R}$, with the unity matrix $\mathbf{1}$ and the usual rotation matrix
\begin{equation}
\mathbf{R} = \left(
\begin{array}{cc}
\cos\phi & \pm\sin\phi \\
\mp \sin\phi & \cos\phi
\end{array}
\right)
\end{equation}
with $\phi = \theta/2$.
For low angles the argument of the exponential becomes
\begin{equation}
i \vec{k}_j^\dagger 
\left(
\begin{array}{cc}
0 & \pm\phi/\epsilon^{1/2} \\
\mp\phi/\epsilon^{1/2} & 0
\end{array}
\right)
\vec{R}.
\end{equation}
Hence the misorientation and $\epsilon$ enter in this limit into the problem only in the combination $\theta/\epsilon^{1/2}$, therefore also the critical misorientation scales as $\theta_c\sim\epsilon^{1/2}$.

As shown in Fig.~\ref{nose},  the range of repulsion is very narrow in PFC for $\epsilon=0.0923$, which corresponds to $\epsilon^{1/2}=0.304$, differing from MD as seen in Fig.~\ref{nose}.  
This result can be interpreted through the ratio of $\gamma_{sl}/\gamma_{gb}$.
A system with a larger ratio prefers grain boundaries to solid-liquid interfaces, narrowing the premelting range, while smaller ratios increase the range of misorientations that premelt. 
With this understanding, in order to obtain the same repulsive range of misorientations as exhibited in MD, we have to adjust $\epsilon$ to decrease the PFC ratio to that of MD. 
Based on the analysis above, the material parameters for MD and PFC for $\epsilon=0.0923$ are summarized in Table  \ref{table2}.
\begin{table}
\begin{center}
\begin{tabular}{c|c|c|clc}
\hline
\hline
Quantity & PFC bcc & AE bcc & MD bcc  \\
\hline
$C_{11}$ (GPa) & 90.0  & 90.0 & 128.0 \\
$C_{12}$ (GPa) & 45.0 & 45.0 & 103.4 \\
$C_{44}$ (GPa) & 45.0 & 45.0 & 63.9  \\
Bulk Modulus (GPa) & \multirow{2}{*}{60.0} & \multirow{2}{*}{60.0} &   \multirow{2}{*}{111.6} \\
$(C_{11}+2C_{12})/3$&  & &\\
$\gamma_{sl}$ ($\mathrm{mJ}/\mathrm{m}^2$) & 160.47 & 144.26 & 177.0  \\
\hline
\hline
\end{tabular}
\caption{Comparison of elastic constants and solid-liquid interfacial energies at the melting point predicted by the PFC model for $\epsilon=0.0923$, AE and MD simulations for bcc Fe \cite{Wuetal2010,WuKarma2007,Sun04,Wuetal2006}. Note that the elastic constants for PFC and AE are the same.
}
\label{table2}
\end{center}
\end{table}
These values are calculated using the procedure put forth in Ref.~\onlinecite{Wuetal2010}, exhibiting the discrepancy between PFC and MD.
In PFC $\gamma_{sl}/B=2.67$ whereas in MD $\gamma_{sl}/B=1.59$, where $B$ denotes the bulk modulus. 
Based on the analysis in Eq.~(\ref{ratioGgamma}) above we can adjust this ratio in PFC by choosing $\epsilon=0.0329$.
In the dimensional units this is achieved by renormalizing $S(q_0)=5.04$ and $C''(q_0)=-17.43\,\mathrm{\AA}^2$ in comparison to the previous values $3.01$ and $-10.4\,\mathrm{\AA}^2$ respectively, thereby leaving their ratio constant, which keeps the value of $\gamma_{sl}$ unchanged, see Eq.~(\ref{gammascaling}).

\begin{table}
\begin{center}
\begin{tabular}{c|c|c|c}
\hline
\hline
Quantity & Original & Rescaled & Unit  \\
\hline
$n_0$ & 0.0765 & 0.0765 & \AA$^{-3}$ \\
$u_s$ & 0.72  & 0.72 &  \\
$q_0$ & 2.985 & 2.985 & \AA$^{-1}$  \\
$a$ & -2.136 & -1.274 & eV\AA$^{3}$  \\
$\lambda$ & 0.291 & 0.488 & eV\AA$^{7}$  \\
$g$ & 9.705 & 258.646 & eV\AA$^{9}$  \\
$\epsilon$ & 0.0923 & 0.0329 &  \\
$S(q_0)$ & 3.01 & 5.04 &  \\
$C''(q_0)$ & -10.40 &  -17.43 & \AA$^{2}$\\
$L$ & $1.968\cdot10^{9}$ & $1.968\cdot10^{9}$ & J/m$^3$ \\
\hline
\hline
\end{tabular}
\caption{Calculated constants and MD input parameters from the MH(SA)$^2$ potential \cite{Mendelevetal2003}. The Original values correspond to $\epsilon=0.0923$ while the Rescaled values correspond to $\epsilon=0.0329$. Calculated constants are derived in Appendix \ref{appenDFT}.
}
\label{table3}
\end{center}
\end{table}

%We know that $\gamma_{sl} \sim \epsilon^{3/2}$ and that $B\sim\epsilon$, thus the ratio of $\gamma_{sl}/B\sim\epsilon^{1/2}$.
%To have the ratio in PFC match that of MD we set $\epsilon=0.0329$.  Changing $\epsilon$ has the additional effect of changing the dimensionless scale of the solid-liquid interfacial energy.  In order to facilitate the comparison between PFC and MD, we are free to change the energy conversion prefactor $\frac{\lambda^2 q_0^5}{g}$ in order to match the relative energy scales.  We see in \cite{Wu07} that for $\epsilon=0.0923$ and $\frac{\lambda^2 q_0^5}{g}= 2.068$ [eV] that $\gamma_{sl,100}=160.47 [$mJ/m$^2]$.  In order to keep the same value of solid-liquid interfacial energy for $\epsilon=0.0329$ we rescale the conversion factor to $\frac{\lambda^2 q_0^5}{g}= 9.719$ [eV].  The rescaling of $\epsilon$ and the conversion prefactor is equivalent to using different values of $S(q_0)$ and $C''(q_0)$, which changes the shape of the liquid structure factor. The new liquid structure factor has $S(q_0)=5.04$ and $C''(q_0)=-17.43$ [\AA$^2$].  Now that the ratio of interfacial energy to elastic moduli is comparable to MD we can then repeat the deterministic simulations for the new value of $\epsilon$. 
\begin{figure}
\centering
\includegraphics[width=0.5\textwidth, angle=0]{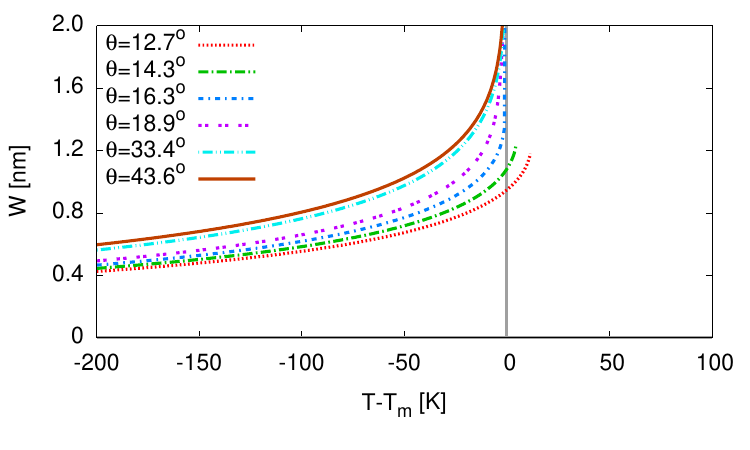}
\caption{(Color online) Liquid layer width as a function of grain boundary width for various misorientations and $\epsilon=0.0329$, obtained from PFC simulations.  There is a transition from attractive to repulsive disjoining potentials at approximately $\theta=15^{\circ}$.}
\label{w3}
\end{figure}

\begin{figure}
\centering
\includegraphics[width=0.5\textwidth, angle=0]{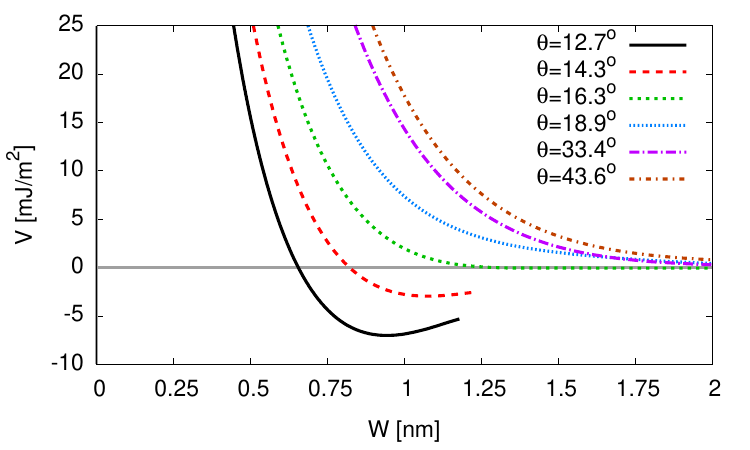}
\caption{(Color online) Disjoining potential as a function of grain boundary width for various misorientations and $\epsilon=0.0329$, obtained from PFC simulations.  There is a transition from attractive to repulsive disjoining potentials at approximately $\theta=15^{\circ}$.}
\label{dispote3}
\end{figure}

\begin{figure}
\begin{center}
\includegraphics[width=9cm]{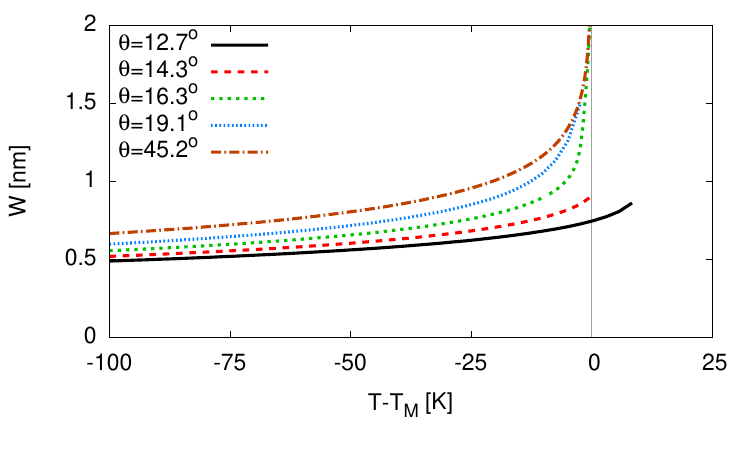}
\caption{(Color online) Melt layer thickness for $\epsilon=0.0329$, as computed from AE. There is a transition from attractive to repulsive disjoining potentials at approximately $\theta=14^{\circ}$.
}
\label{glsmalleps2}
\end{center}
\end{figure}

\begin{figure}
\begin{center}
\includegraphics[width=9cm]{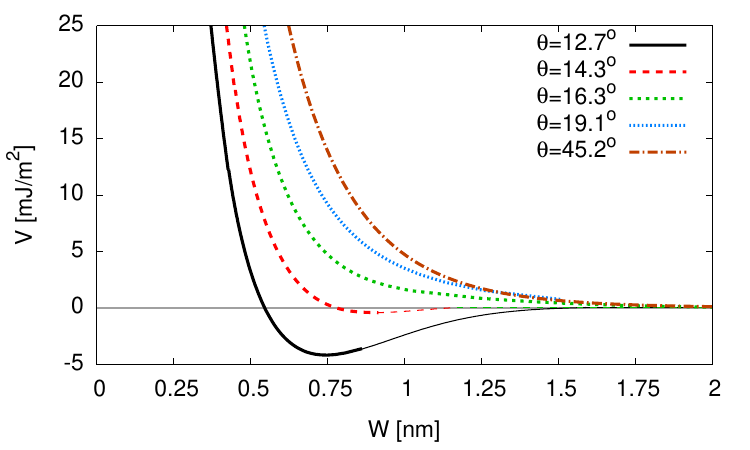}
\caption{(Color online) Disjoining potential for $\epsilon=0.0329$, as computed from AE. There is a transition from attractive to repulsive disjoining potentials at approximately $\theta=14^{\circ}$.
}
\label{glsmalleps1}
\end{center}
\end{figure}

\begin{figure}
\begin{center}
\includegraphics[width=9cm]{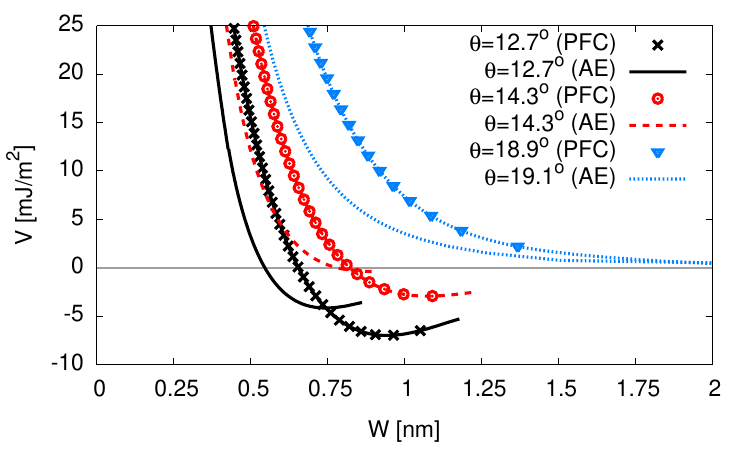}
\caption{(Color online) Disjoining potential comparison between PFC and AE for $\epsilon=0.0329$ for selected misorientations.
}
\label{glsmalleps3}
\end{center}
\end{figure}

\begin{figure}
\begin{center}
\includegraphics[width=9cm]{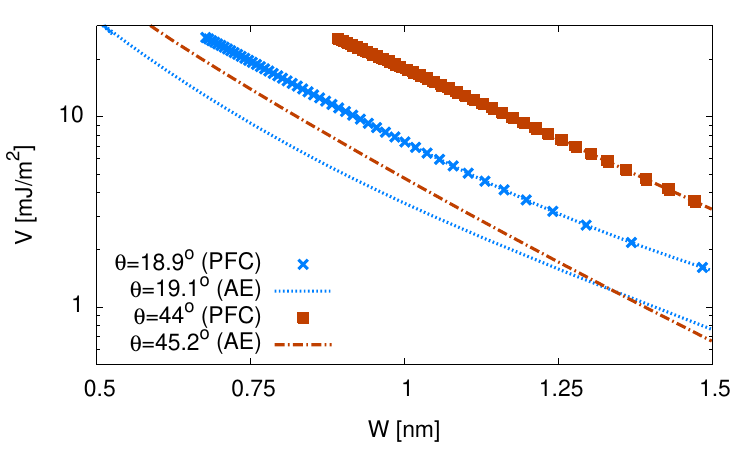}
\caption{(Color online) Disjoining potential comparison for $\epsilon=0.0329$ for selected misorientations. The exponential decay for the corresponding angles is similar for PFC and AE.
}
\label{glsmalleps4}
\end{center}
\end{figure}

As seen in Fig.~\ref{dispote3}, for the new value of $\epsilon$ the critical angle has shifted to a value of $\theta\approx 15^{\circ}$.
The corresponding results for AE are shown in Figs.~\ref{glsmalleps2} and \ref{glsmalleps1}.
Since the value of $\epsilon$ is smaller here, the agreement between AE and PFC is significantly better than for the higher $\epsilon$ value.
A direct comparison is shown in Fig.~\ref{glsmalleps3}.
The logarithmic plot Fig.~\ref{glsmalleps4} shows that the decay ranges are very much comparable for AE and PFC.
As demonstrated in Ref.~\onlinecite{ShiftedCrystalPaper}, the decay length also agrees with analytical predictions.

%\begin{figure}
%\centering
%\includegraphics[width=0.5\textwidth, angle=0]{PFCGL/pfcampmdboth.pdf}
%\caption{A comparison of disjoining potentials between PFC, GL, and MD for a $\theta\approx44^{\circ}$ misorientation, with PFC results for both $\epsilon=0.0923$ and $\epsilon=0.0329$.}
%\label{compare9a3}
%\end{figure}

\begin{figure}
\centering
\includegraphics[width=0.5\textwidth, angle=0]{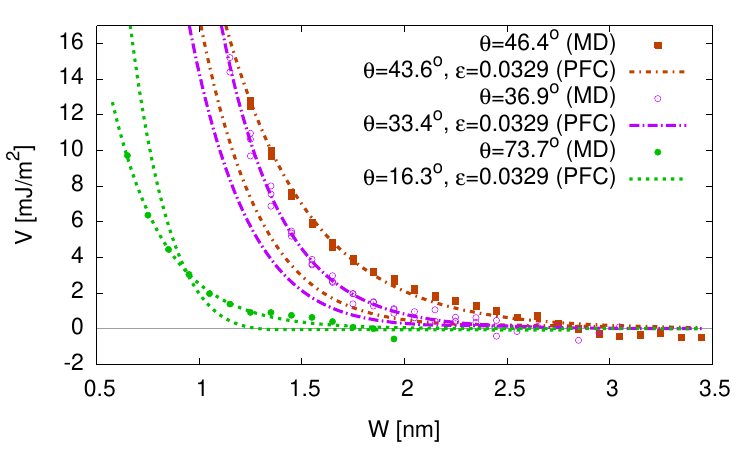}
\caption{(Color online) Three different MD disjoining potentials.
The points are the simulation data, the lines are exponential fits.
%{\bf Ari - The tail of the $46.4^\circ$ degree does fall below 0, but this does provide the best fit to the exponential. I'm not sure what the MD procedure is in this case.  Do we require that the tail goes to 0?, or is the exponential fit more important?}
The results are compared to PFC data for the rescaled value of $\epsilon=0.0329$.
}
\label{Mdall}
\end{figure}

In Fig.~\ref{Mdall} we also present three different MD disjoining potentials, over a similar range of angles as PFC and AE.
We see that the range of of the interaction for PFC and AE is now comparable to the range of interaction in MD. 

%%%%%%%%%%%%%%%%%%%%%%%%%%%%%%%%%%%%%%%%%%%%%%%%%%%%%%%%%%%
\subsection{Fluctuation effects }
\label{fluctuations}

As seen in Fig. \ref{compare9} there is a profound difference in the range of the disjoining potential for MD and PFC with $\epsilon=0.0923$.  Above, we attempted to fix this difference through a rescaling of the parameter $\epsilon$.  Another possible route is to look at the role of fluctuations in PFC.  
Intuitively spoken, the underlying idea is that the thermal fluctuations lead to a broadening of the melt layer due to the asymmetric nature of the disjoining potential $V(W)$.
We quantitatively incorporate fluctuations into our calculation of the disjoining potential for one high angle boundary. As discussed above in Section \ref{cutoff} we choose to cutoff the noise on the scale of one unit cell.
\begin{figure}
\centering
\includegraphics[width=0.5\textwidth, angle=0]{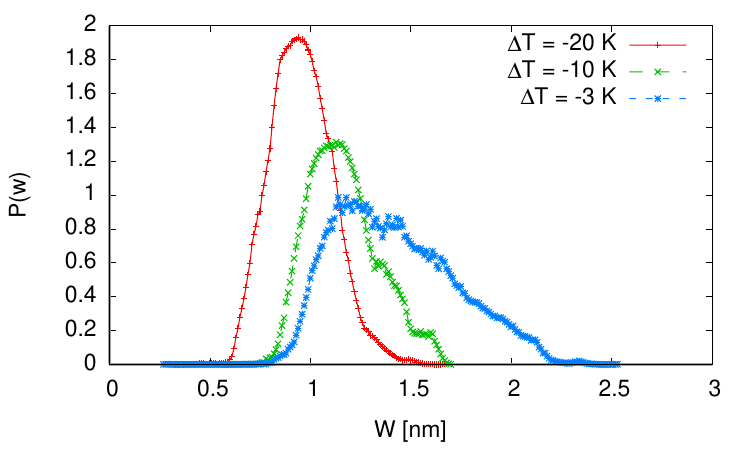}
\caption{(Color online) Histograms of PFC with noise grain-boundary liquid layer width as a function of three different undercoolings for a misorientation of $43.6^{\circ}$, $\epsilon=0.0923$, and a cutoff of the noise at the scale of one unit cell $(\lambda_{min}=a$).}
\label{hk1}
\end{figure}
In Fig. \ref{hk1} histograms of the width distributions for three different undercoolings for a misorientation of $\theta=43.6^{\circ}$ are shown.  Using the procedure presented above in Section \ref{thermfluc}, we can extract the disjoining potential.
Fig. \ref{d9n} shows a comparison of MD disjoining potentials, PFC without noise, and noise based PFC for two different values of $\lambda_{min}$.  Akin to renormalizing $\epsilon$, the fluctuation effects clearly increase the range of the disjoining potential.
However, the strength of this renormalization is dependent on $\lambda_{min}$.  

The introduction of fluctuations has the additional effect of renormalizing the equilibrium chemical potential ($\mu_{eq}$).  Without noise $\epsilon=0.0923$ and $\mu_{eq}\approx -0.19609$.  The renormalized value of the equilibrium chemical potential is calculated by running cyrstal-liquid coexistence simulations, and we determine that for $\lambda_{min} = 1a$, $\mu_{eq}\approx-0.18721$ and for $\lambda_{min} = 2a$, $\mu_{eq}\approx-0.19565$, where $a$ is one lattice spacing.

\begin{figure}
\centering
\includegraphics[width=0.5\textwidth, angle=0]{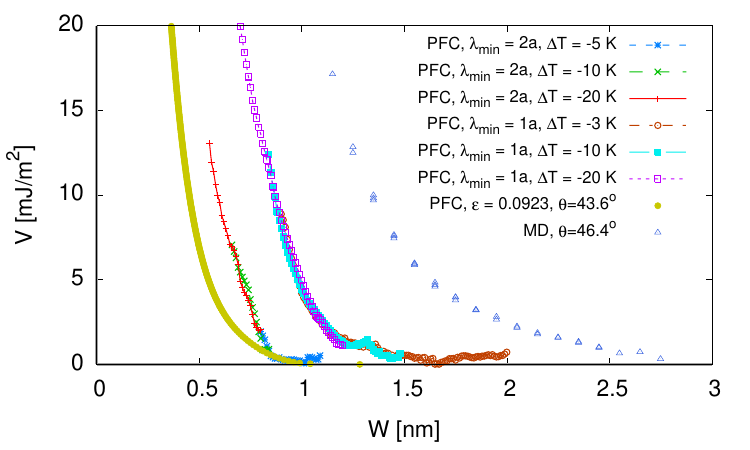}
\caption{(Color online) A comparison of PFC disjoining potentials with and without noise with an MD disjoining potential.  The PFC results are for $\theta=43.6^{\circ}$ and $\epsilon=0.0923$ for two different values of $\lambda_{min}$, while the MD results are for $\theta=46.4^{\circ}$ and are for three different undercoolings. $a$ is the length of one lattice spacing.
}
\label{d9n}
\end{figure}

%%%%%%%%%%%%%%%%%%%%%%%%%%%%%%%%%%%%%%%%%%%%%%%%%%%%%%%%%%%
\subsection{Summary}

\begin{figure}
\begin{center}
\includegraphics[width=9cm]{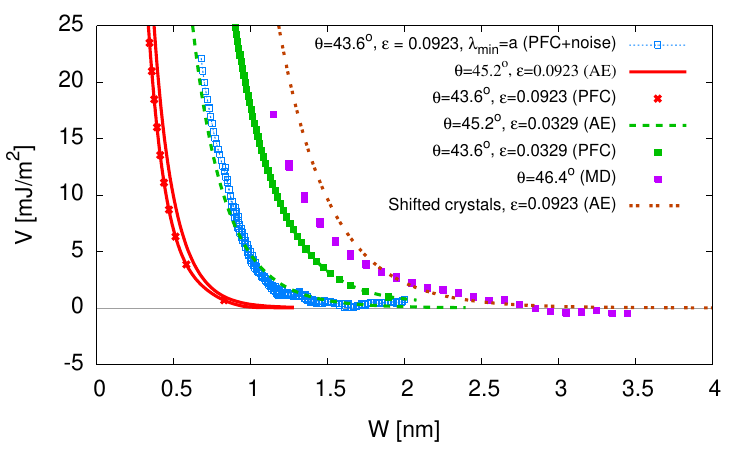}
\caption{
(Color online)
Comparison of results for PFC and AE for the two different values of $\epsilon$ with the PFC results with noise and the MD data.
Additionally, the AE data for two crystals, which do not have a misorientation but which are just shifted against each other by half a lattice unit is shown.
This curve is for a normal direction $21.8^\circ$ off [100], which is the same as for the symmetric tilt grain boundaries, but here both grains are rotated in the same direction.
%For comparison, an estimate for the nonretarded dispersion forces is shown for a Hamaker constant of $H=10^{-21}\,\mathrm{J}$.
}
\label{sum1}
\end{center}
\end{figure}

\begin{figure}
\begin{center}
\includegraphics[width=9cm]{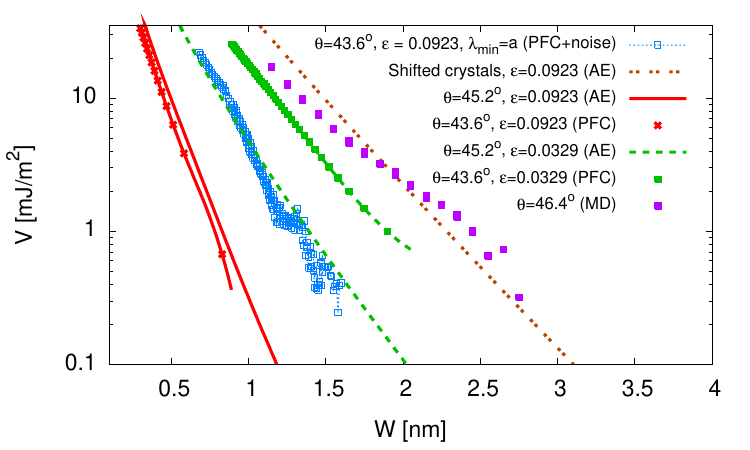}
\caption{
(Color online)
The same as Fig.~\ref{sum1}, but on a logarithmic scale, to show the interaction range for the different models.
}
\label{sum2}
\end{center}
\end{figure}

All the essential data are summarized in Figs.~\ref{sum1} and \ref{sum2}.
The plots show that the rescaling of $\epsilon$ has a similar effect as the inclusion of thermal fluctuations, bringing the PFC and the MD results closer together.
The logarithmic plot shows that the interaction is indeed short ranged and decays exponentially for all simulation techniques.
Nevertheless, there is still a significant difference in the slopes, i.e.~the range of the interaction, between MD and the deterministic calculations.
In comparison, the simulations with fluctuations exhibit a very similar behavior for short distances, because there the thermal fluctuations cannot compete with the overall strength of the repulsion.
At larger grain separations, $W>1.5\,\mathrm{nm}$, the data scatter becomes larger, and the change of slope may also be due to insufficient knowledge of the solid-liquid interfacial energy.
Nevertheless, there is a clear tendency, that (i) by better matching of the elastic constants, and (ii) incorporation of thermal fluctuations a better agreement of continuum methods and MD simulation results can be achieved.

Finally, the plot also demonstrates that at on the relevant lengthscales the structural short range forces are substantially larger than London forces.
Non-retarded dispersion forces between planar interfaces are asymptotically described by the interaction potential
\begin{equation}
V_{d} \simeq \frac{H}{12\pi W^2},
\end{equation}
with the Hamaker constant $H$.
For a characteristic value $H=10^{-21}\,\mathrm{J}$ (see Ref.~\onlinecite{Pluis89}) the dispersion forces are 
%included in Figs.~\ref{sum1} and \ref{sum2}
of the order $V_{d}\sim 10^{-2}\,\mathrm{mJ/m^2}$ at distances $W\sim 1\,\mathrm{nm}$, showing that their contribution is negligible.
We also note that additional repulsive entropic forces are completely negligible in comparison to the short-range structural forces computed here, see also the discussion in Ref.~\onlinecite{Hoytetal2009}.

%%%%%%%%%%%%%%%%%%%%%%%%%%%%%%%%%%%%%%%%%%%%%%%%%%%%
\section{Grain boundary shearing}
\label{shearing::sec}

In this section we use PFC to explore the relationship between premelting  and the response of grain boundaries to shear stress. Dry symmetric tilt grain boundaries, or in PFC boundaries which are sufficiently undercooled, are known to follow a strict geometrical relationship between the velocity of the applied shear parallel to the boundary ($v_{||}$) and its motion normal to itself ($v_n$) \cite{CahnTaylor2004,Cahnetal2006}. This geometric model of  coupling predicts two different branches of motion, $\beta_{\langle100\rangle}$ and $\beta_{\langle110\rangle}$ where
\begin{eqnarray}
v_{||}&=&\beta v_n,\\
\beta_{\langle100\rangle}&=&2\tan \left(\frac{\theta}{2}\right),\\
\beta_{\langle110\rangle}&=&-2\tan\left(\frac{\pi}{4}-\frac{\theta}{2}\right),
\end{eqnarray}
where $\theta$ is the angle of total misorientation.  Which coupling branch is preferred depends on the Burgers vector content of the grain boundary. For bcc low angle misorientaions with normal near $(100)$, $\vec{b} = [100]a$ where $a$ is the lattice spacing. These boundaries couple with $\beta = \beta_{\langle100\rangle}$. For low angle misorientations with normal near $(110)$, $\vec{b} = [110]a$ and $\beta = \beta_{\langle110\rangle}$. For misorientations where the dislocations overlap, either coupling branch is accessible.  In order to confirm the coupling branches for PFC we shear a bcc crystal with a symmetric tilt grain boundary using the dynamics presented in Eq.~(\ref{dotdyn}). These dynamics quickly relax elastic strain allowing us to shear at a faster velocity without deforming the crystal.  To shear the crystal we add a term to the free-energy functional of the form,
\begin{eqnarray}
{F}'&=F+\int d\vec{V}g(x)(\psi-\psi_0)^2=\int d\vec{V}f',
\end{eqnarray}
where $g(x)$ is a normalized Gaussian of the form
\[
g(x) = \frac{1}{\sqrt{2\pi\sigma^2}}e^{-\frac{1}{2}(\frac{x-x_c}{\sigma})^2}, 
\]
with $\sigma=a$ (one unit cell) and where
\begin{equation}
\psi_0=\psi_f(x,y\pm vt,z).
\end{equation}
The expression $\psi_f$ is calculated using Eq.~(\ref{RLV}).  The added term to the free-energy effectively drags a strip of the crystal with a constant velocity of $\pm v$.  In our systems we place one Gaussian strip to the right of the grain boundary shearing downwards, and one strip to the left of the grain boundary shearing upwards.  We perform simulations for a full range of misorientations $\theta$ with renormalized $\epsilon=0.0329$ in order to better compare to MD results and at an undercooling which ensures the physical contact of the two sides of the grain boundary (meaning a negligible liquid layer), and a net shearing velocity of $v_{||} = 0.0005$.  In Fig. \ref{couplef} we see that for low temperature, PFC does reproduce the predicted ideal coupling, switching between the two branches at $\theta=45^{\circ}$.

\begin{figure}
\centering
\includegraphics[width=0.45\textwidth, angle=0]{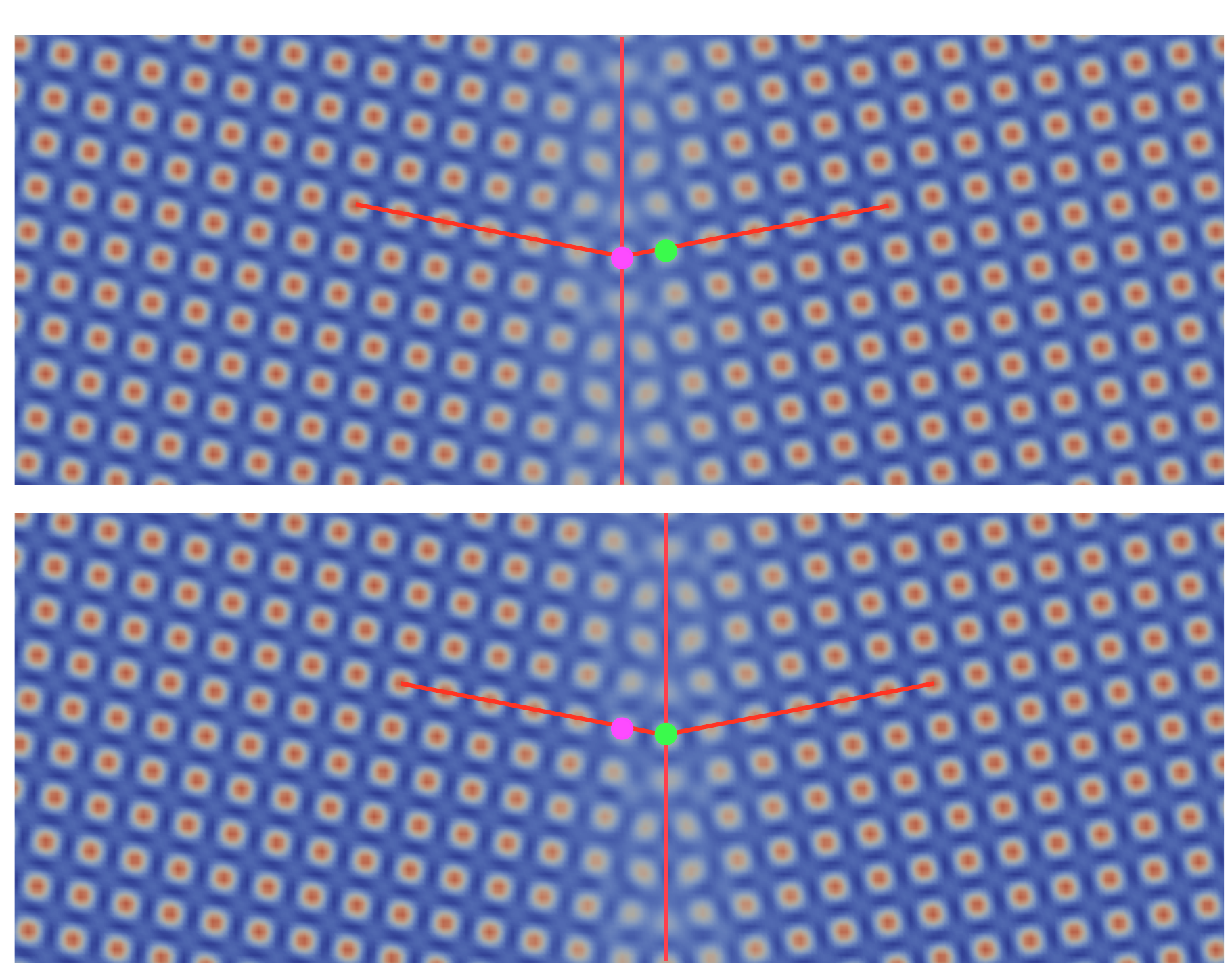}
\caption{ (Color online) An example of $\beta_{\langle100\rangle}$ symmetric tilt grain boundary coupling.  The grain on the right is being sheared downwards with velocity $v=-0.00025$, while the grain on the left is being sheared upwards with velocity $v=0.00025$.  The green atom starts in the right crystal, but due to the rightward motion of the boundary becomes attached to both crystals.  It will eventually follow the pink atom and join a (100) plane of the left crystal.}
\label{bccsf}
\end{figure}

\begin{figure}
\centering
\includegraphics[width=0.5\textwidth, angle=0]{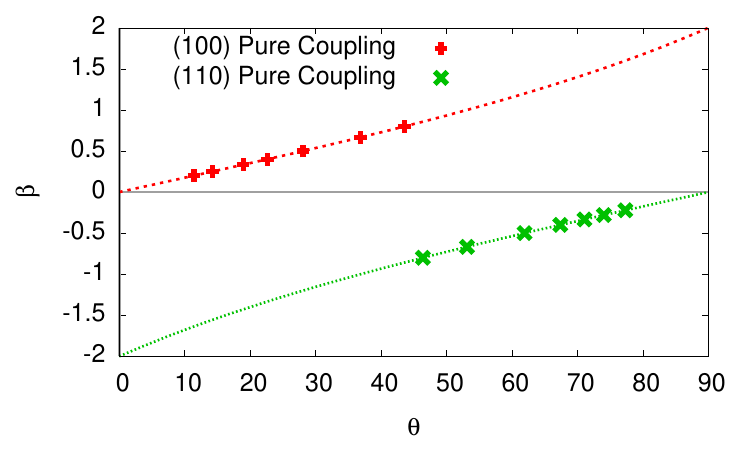}
\caption{(Color online) Observed $\beta$ as a function of misorientation $\theta$ for $\epsilon=0.0329$, and the ideal $\beta$ factors for $(100)$ and $(110)$ coupling.  These simulations were conducted at $T/T_M\approx 0.7$, and a net shearing velocity of $v_{||} = 0.0005$.}
\label{couplef}
\end{figure}
A recent study \cite{Olmstedetal2011} explored the existence of an alternative grain boundary structure for low angle misorientations with normal near $(110)$. This ``unpaired" structure splits the $[022]a$ dislocations into pairs of $[\bar{1}11]a/2$ and $[111]a/2$ dislocations. Simulations performed with this unpaired grain boundary structure exhibited the same observed coupling factor as the paired grain boundary. This is easily understood as the net Burgers vector content of the two boundaries is the same. It is also possible at temperatures closer to $T_M$, where the atoms on either side of the grain boundary are no longer in immediate contact, to observe behaviors other then pure coupling. An MD study of Ni (Ref.~\onlinecite{Cahnetal2006}) observed pure sliding for some misorientations as the melting temperature was approached from below.  In order to delineate these regions we perform a series of simulations over the full range of misorientations, and over a temperature range extending from an undercooling where we observe pure coupling at all $\theta$ up to temperatures 1 K below $T_M$.

\begin{figure}
\centering
\includegraphics[width=0.5\textwidth, angle=0]{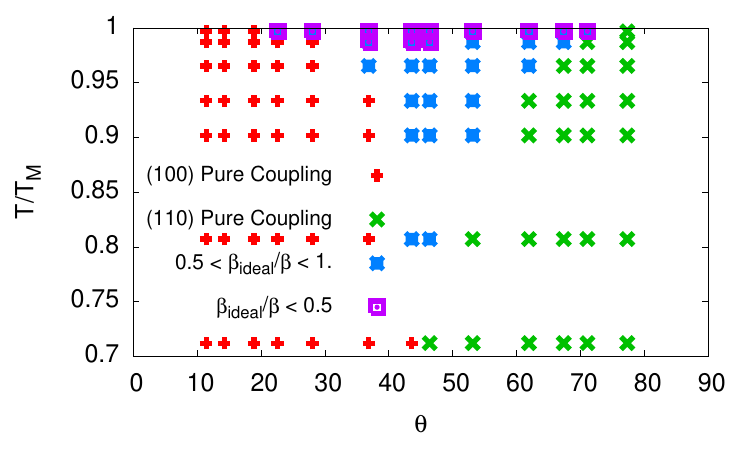}
\caption{ (Color online) Observed $\beta$ as a function of misorientation $\theta$  and homologous temperature for $\epsilon=0.0329$. We shear with a net velocity $v_{||}=0.0005$.
}
\label{maphomol}
\end{figure}
We see in Fig. \ref{maphomol} that for temperatures closer to the melting temperature we observe coupling modes other than ideal coupling.  There is a "v" shaped region where the observed $\beta$ is larger than $\beta_{ideal}$.

\begin{figure}
\centering
\includegraphics[width=0.5\textwidth, angle=0]{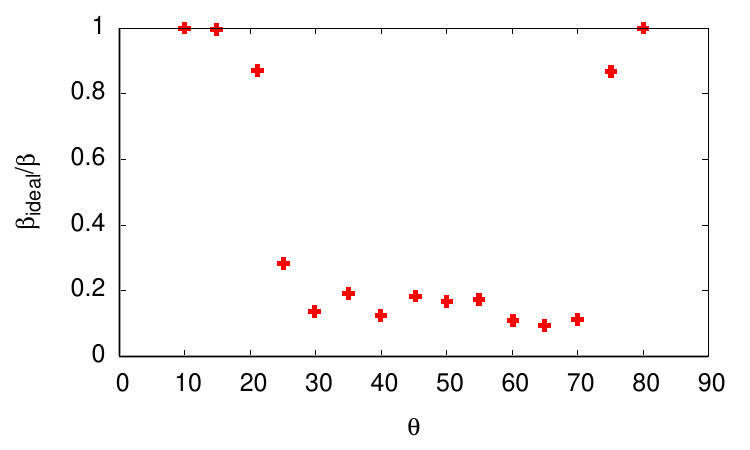}
\caption{ (Color online) The ratio of $\beta_{ideal}$ to the observed $\beta$ as a function of misorientation for an undercooling of $0.46$ K corresponding to the top row of Fig.~\ref{maphomol}. Simulations are performed with $v_{||}=0.0005$ and $\epsilon=0.0329$. There is a transition from ideal coupling to something approaching pure sliding as a function of misorientation.
}
\label{hightcouple}
\end{figure}
In Fig.~\ref{hightcouple} we examine the observed coupling factor close to $T_M$. For low angle misorientations, where the liquid layer width is smaller, we still observe pure coupling. High angle misorientations however, have coupling factors much larger than ideal.

%%%%%%%%%%%%%%%%%%%%%%%%%%%%%%%%%%%%%%%%%%%%%%%%%%%%%%%%%%%

\section{Conclusions}
\label{summary::sec}

We have used the PFC and AE methods to investigate the equilibrium premelting and nonequilibrium shearing behaviors of $[001]$ symmetric tilt grain boundaries (GBs) at high homologous temperature in classical models of bcc Fe, and also compared our equilibrium results  to MD simulations. 

At a qualitative level, our findings are consistent with the findings of previous PFC \cite{Mellenthinetal2008} and MD \cite{Hoytetal2009,Fensinetal2010} studies. We find that the disjoining potential can be either repulsive over an intermediate range of misorientation ($\theta_{\rm min}<\theta<\theta_{\rm max}$), or exhibit an attractive minimum outside this range ($\theta<\theta_{\rm min}$ or $\theta_{\rm max}<\theta<90^\circ$). At a more quantitative level, we find that the repulsive range of misorientation is not well predicted by the PFC model for the value of $\epsilon$ matched to liquid structure factor properties. The range predicted by PFC is too narrow compared to MD \cite{Olmstedetal2011}. This disagreement stems from the fact that the PFC model is too simplified to reproduce, for the same value of $\epsilon$, the correct liquid structure factor properties, the correct solid-liquid interfacial free-energy, and the correct elastic moduli. However, if $\epsilon$ is chosen to have a lower value so as to match the ratio of bulk modulus and $\gamma_{sl}$ in the MD simulations, the PFC model predicts accurately the range of misorientation where GB exhibit a diverging layer width at the melting point. Since the elastic moduli are a major determinant of the GB free-energy at the melting point, this result indicates that elastic properties and the solid-liquid interfacial free-energy play a dominant role in setting the range of misorientation where GBs premelt. In the AE framework, the range of misorientation where GBs remain dry, with a minimum in the disjoining potential, is predicted to shrink to zero in the small $\epsilon$ limit with $\theta_{\rm min}$ and $90^\circ-\theta_{\rm max}$ both scaling as $\sim \epsilon^{1/2}$. 

Furthermore, we have found that with the lower value of $\epsilon$, which predicts the correct premelted range of misorientation, the PFC model reproduces the expected GB shearing behavior as a function of homologous temperature and misorientation \cite{Cahnetal2006}.  
Namely, for low enough homologous temperature, GBs exhibit pure coupled motion to a shear stress with a discontinuous change of the coupling factor as a function of $\theta$ that reflects a transition between two coupling modes. In contrast, for high enough homologous temperature ($T/T_M\ge 0.8$ in the PFC model), partial sliding occurs, thereby causing the coupling factor $\beta$ to be lower than its ideal value. The range of misorientation where $\beta<\beta_{ideal}$ increases with temperature but remains finite even at the melting point since low angle GBs with well separated dislocations exhibit pure coupled motion even when the dislocation cores are partially melted. 

Those results support the view that, for symmetric tilt GBs, partial sliding is primarily due to the formation of a premelted layer at high homologous temperature. However a recent combined PFC and MD study \cite{Trauttetal2012} has shown that asymmetrical tilt GBs can exhibit partial sliding even for low temperatures where such a layer is absent. Therefore further studies remain needed to obtain a more complete understanding of GB sliding mechanisms as a function of temperature and GB bi-crystallography.  

%%%%%%%%%%%%%%%%%%%%%%%%%%%%%%%%%%%%%%%%%%%%%%%%%%%%%%%%%%%

\begin{acknowledgments}

This research was supported by grants from the US Department of Energy, Office of Basic Energy Sciences, under Contracts No. DE-FG02-07ER46400 (AA and AK) and DE-FG02-06ER46282 (DB and MA), and the Computational Materials Science Network program.  The MD simulations made use of resources at the National Research Scientific Computing Center (DOE Office of Science Contract No. DE-AC02-05CH11231). RS acknowledges support by the grant SP 1153/1-3 of the German Research Foundation.

\end{acknowledgments}

%%%%%%%%%%%%%%%%%%%%%%%%%%%%%%%%%%%%%%%%%%%%%%%%%%%%%%%%%%%

\appendix

%%%%%%%%%%%%%%%%%%%%%%%%%%%%%%
\section{Integration schemes}

%%%%%%%%%%%%%%%%%%%%%%%%%%%%%%
\subsection{PFC with noise}
\label{dynamicsA}
The evolution equations for non-conserved dynamics with noise are given in Eq.~(\ref{fulldyn}). Due to the high order derivative terms, it is computationally more tractable to conduct the simulations in reciprocal space. We apply a Fourier transform as in the deterministic dynamics in the Appendix of Ref.~\onlinecite{Mellenthinetal2008},
\begin{eqnarray}
\partial_t\tilde{\psi}_k&=&\hat{L}_k\tilde{\psi}_k+\tilde{g}_k,\\
\hat{L}_k&=&(\epsilon-1)+2k^2-k^4,\\
\tilde{g}_k&=&-\int d\vec{r} (\psi^3)e^{i \vec{k}\cdot\vec{r}}+\tilde{\mu}_k+\tilde{\eta}_k\nonumber\\
&=&\tilde{f}_k+\tilde{\eta}_k,\\
\end{eqnarray}
We can now proceed as in Ref.~\onlinecite{Mellenthinetal2008}
\begin{eqnarray}
\tilde{\psi}_k &=& u(t)e^{\hat{L}_kt},\\
\partial_t \tilde{\psi}_k&=&\hat{L}_k e^{\hat{L}_kt}u(t)+\tilde{g}_k,\\
\partial_t u(t)&=&e^{-\hat{L}_k t}\tilde{g}_k.
\end{eqnarray}
Integrating $u(t)$ in time gives us the expression,
\begin{equation}
u(t+\Delta t)-u(t)=\int_t^{t+\Delta t} dt' \exp(-\hat{L}_k t)\tilde{g}_k(t'),
\end{equation}
where $\tilde{f}_k(t')$ can be expanded around $t'=t$.  Since the term $\tilde{\eta}_k$ is an instantaneous value over the interval $\Delta t$ it can be treated as a constant during integration.  Our final expression for the dynamics in a continuous case is then
\begin{eqnarray}
\tilde{\psi}_k(t+\Delta t)&=&e^{\hat{L}_k \Delta t}\tilde{\psi}_k(t) + e^{\hat{L}_k (t+\Delta t)}\int_t^{t+\Delta t} dt' e^{-\hat{L}_k t'}\nonumber\\
&\times &\left[ \tilde{f}_k(t) +\frac{\tilde{f}_k(t)-\tilde{f}_k(t-\Delta t)}{\Delta t}(t'-t) +\tilde{\eta}_k\right]\nonumber\\
&=&e^{\hat{L}_k \Delta t}\tilde{\psi}_k(t)+\frac{\tilde{f}_k(t)}{\hat{L}_k}(e^{\hat{L}_k\Delta t}-1)\nonumber\\
&+&\frac{\tilde{f}_k(t)-\tilde{f}_k(t-\Delta t)}{\Delta t \hat{L}_k^2}(e^{\hat{L}_k\Delta t}-1-\Delta t\hat{L}_k)\nonumber\\
&-&\frac{1-e^{\hat{L}_k \Delta t}}{\hat{L}_k}\tilde{\eta}_k.
\end{eqnarray}
Special care must be taken to properly normalize the noise in spectral space. In real space the average noise and two point correlation function of the noise are
\begin{eqnarray}
\langle\eta(\vec{r},t)\rangle&=&0, \label{noise::eq1}\\
\langle\eta(\vec{r},t)\eta'(\vec{r}',t')\rangle&=&\frac{2k_BTg}{\lambda^2q_0^5}\delta(\vec{r}-\vec{r}')\delta(t-t'). \label{noise::eq2}
\end{eqnarray}
The last equation includes the strength of the Gaussian noise ($2k_BT$) converted into PFC dimensionless units.
While these equations accurately describe the addition of noise for a continuous system in real space, care must be taken to normalize the noise properly in a discretized spectral space, specifically for the fast Fourier transform which we use in this study.
We start for simplicity in one dimension where
\begin{equation}
\langle\eta(x,t)\eta'(x',t')\rangle=\frac{2k_BTg}{\lambda^2q_0^5}\delta(x-x')\delta(t-t').
\end{equation}
For a discretized system we write
\begin{equation}
\langle \eta_j^n\eta_l^m\rangle = \frac{2k_BTg}{\lambda^2q_0^5}\frac{\delta_{jl}}{\Delta x}\frac{\delta_{nm}}{\Delta t},
\end{equation}
where $j,l,m,n$ are integers, $x(x')=j(l)\Delta x$, and $t(t') = n(m)\Delta t$.  We then define the discretized Fourier transform of the noise to be
\begin{eqnarray}
\tilde{\eta}_r^n = \sum_{j=0}^{N-1} e^{-2\pi i j r/N} \eta_j^n,\nonumber\\
\tilde{\eta}_s^m = \sum_{l=0}^{N-1} e^{-2\pi i l s/N} \eta_l^m,
\end{eqnarray}
where $N$ is the number of points in the system and $r(s)$ is an integer that takes on all values from $-N/2+1$ to $N/2$ inclusive. $r$ can be related to the frequency through the formula $k = 2\pi r/N\Delta x$. The noise in Fourier space must be normalized separately for both real and imaginary parts.
\begin{equation}
\tilde{\eta}_r^n=\tilde{\eta}^{n,R}_r-i\tilde{\eta}^{n,I}_r,
\end{equation}
where
\begin{eqnarray}
&&\tilde{\eta}^{n,R}_r= \sum_{j=0}^{N-1} \cos \frac{2\pi j r}{N}\eta^n_j,\\
&&\tilde{\eta}^{n,I}_r=\sum_{j=0}^{N-1} \sin \frac{2\pi j r}{N}\eta^n_j,\\
&&\langle\tilde{\eta}_r^{n}\tilde{\eta}_{s}^m\rangle=\nonumber\\
&&\langle\tilde{\eta}^{n,R}_r\tilde{\eta}^{m,R}_{s}\rangle-\langle\tilde{\eta}^{n,I}_r\tilde{\eta}^{m,I}_{s}\rangle\nonumber\\
&&-i\langle\tilde{\eta}^{n,R}_r\tilde{\eta}^{m,I}_{s}\rangle-i\langle\tilde{\eta}^{n,I}_r\tilde{\eta}^{m,R}_{s}\rangle.
\end{eqnarray}
It can easily be seen that the two crossterms do not contribute to the overall noise,
\begin{eqnarray}
&&\langle\tilde{\eta}^{n,R}_s\tilde{\eta}^{m,I}_{s}\rangle\nonumber\\
=&&\frac{2k_BTg}{\lambda^2q_0^5}\frac{\delta_{nm}}{\Delta t}\sum_{j,l=0}^{N-1}\frac{\delta_{jl}}{\Delta x}\cos \frac{2\pi j r}{N}\sin\frac{2\pi l s}{N},\\
=&&\frac{2k_BTg}{\lambda^2q_0^5}\frac{\delta_{nm}}{\Delta t\Delta x}\sum_{j=0}^{N-1}\cos\frac{2\pi j r}{N}\sin\frac{2\pi j s}{N},
\end{eqnarray}
which is equal to zero for all integers $r$ and $s$.
The other two terms do contribute to the noise,
\begin{eqnarray}
&&\langle\tilde{\eta}^{n,R}_r\tilde{\eta}^{m,R}_{s}\rangle\nonumber\\
=&&\frac{2k_BTg}{\lambda^2q_0^5}\frac{\delta_{nm}}{\Delta t}\sum_{j,l=0}^{N-1}\frac{\delta_{jl}}{\Delta x}\cos\frac{2\pi j r}{N}\cos\frac{2\pi l s}{N},\\
=&&\frac{2k_BTg}{\lambda^2q_0^5}\frac{\delta_{mn}}{\Delta t\Delta x}\sum_{j=0}^{N-1}\cos \frac{2\pi j r}{N}\cos\frac{2\pi j s}{N},\\
=&&\frac{2k_BTg\delta_{mn}}{\lambda^2q_0^5\Delta x\Delta t}\delta_{rs}\left (\frac{N}{2}+\frac{N}{2}(\delta_{r0}+\delta_{rN/2})\right),
\end{eqnarray}
and
\begin{eqnarray}
&&\langle\tilde{\eta}^{n,I}_r\tilde{\eta}^{m,I}_{s}\rangle\nonumber\\
=&&\frac{2k_BTg}{\lambda^2q_0^5}\frac{\delta_{mn}}{\Delta t}\sum_{j,l=0}^{N-1}\frac{\delta_{jl}}{\Delta x}\sin\frac{2\pi j r}{N}\sin\frac{2\pi l s}{N},\\
=&&\frac{2k_BTg}{\lambda^2q_0^5}\frac{\delta_{nm}}{\Delta t\Delta x}\sum_{j=0}^{N-1}\sin\frac{2\pi j r}{N}\sin\frac{2\pi j s}{N},\\
=&&\frac{2k_BTg\delta_{mn}}{\lambda^2q_0^5\Delta x\Delta t}\delta_{rs}\left (\frac{N}{2}-\frac{N}{2}(\delta_{r0}+\delta_{rN/2})\right).
\end{eqnarray}
As can be seen both the imaginary and real terms contribute equally for $r,s\neq 0$, but the $r,s=0$ and $r,s=N/2$ terms have the full magnitude only on the real part of the dynamics, and are zero for the imaginary part.  The logic extends easily to three dimensions with $N/\Delta x$ replaced by $(N_xN_yN_z)/(\Delta x\Delta y\Delta z)$. As presented in section \ref{cutoff}, in order to ensure that the noise based contribution to the free-energy is less than the free-energy of a noiseless system we must cutoff the noise at a length scale close to one lattice spacing. We choose the following form for our cutoff

\begin{eqnarray}
\tilde{\eta}^{n,R}_r&=&G\sqrt{\frac{\Gamma}{\Delta x\Delta t}\left (\frac{N}{2}+\frac{N}{2}(\delta_{r0}+\delta_{rN/2})\right) }  \times \nonumber\\
&&\left( \frac{1}{2}+ \frac{1}{2}\tanh\frac{\lambda-\lambda_{min}}{\xi} \right),\nonumber\\
\\
\tilde{\eta}^{n,I}_r&=&G\sqrt{\frac{\Gamma}{\Delta x\Delta t}\left (\frac{N}{2}-\frac{N}{2}(\delta_{r0}+\delta_{rN/2})\right)} \times\nonumber\\
&&\left( \frac{1}{2}+ \frac{1}{2}\tanh\frac{\lambda-\lambda_{min}}{\xi} \right),\nonumber\\
\end{eqnarray}
where $\xi$ controls the width of the cutoff, $\lambda_{min}$ is the minimum noise wavelength, $\Gamma=2k_BTg/(\lambda^2q_0^5)$, and $G$ are Gaussian random numbers with a standard deviation of one. 

%%%%%%%%%%%%%%%%%%%%%%%%%%%%%%%%%%%
\subsection{Wave Dynamics}
\label{dynamicsB}

Here we present the algorithm used to implement the modified PFC dynamics introduced in Ref.~\onlinecite{Stefanovicetal2009}.
We repeat the free-energy and dynamical equation for locally conserved dynamics
\begin{eqnarray}
&&F=\int d\vec{r} \left[\frac{\psi}{2}(-\epsilon+(\nabla^2+1)^2)\psi+\frac{\psi^4}{4}+\frac{\dot{\psi}^2}{2}\right],\nonumber\\
\\
&&\partial_{tt}\psi+\alpha\partial_t \psi = \nabla^2 \frac{\delta F}{\delta \psi}.
\end{eqnarray}
As in Ref.~\onlinecite{Mellenthinetal2008} we avoid calculating the gradient terms in real space by solving the equation of motion in Fourier space.  We define a Fourier transform
\begin{equation}
\tilde{\psi}_k = \int d\vec{r} e^{i\vec{k}\cdot \vec{r}}\psi.
\end{equation}
We multiply both sides of the dynamical equation by $\exp(i\vec{k}\cdot \vec{r})$ and integrate.  The equation of motion in Fourier space is now
\begin{equation} \label{Apdyn}
\partial_{tt}\tilde{\psi}_k+\alpha\partial_t \tilde{\psi}_k = \hat{L}_k\tilde{\psi}_k+\tilde{f}_k,
\end{equation}
where
\begin{eqnarray}
\hat{L}_k&=&-k^{6}+2k^4+(\epsilon-1)k^2,\\
\tilde{f}_k &=& \int d\vec{r} e^{i\vec{k}\cdot \vec{r}} (\nabla^2\psi^3).
\end{eqnarray}
Following the methodology of Section 3 of Chapter II of Ref.~\onlinecite{Chand43}, we now choose to rewrite the dynamics as
\begin{eqnarray}
\dot{\tilde{\psi}}_k&=&\tilde{u}_k,\\
\dot{\tilde{u}}_k &=& -\alpha \tilde{u}_k +\hat{L}_k\tilde{\psi}_k + \tilde{f}_k.
\end{eqnarray}
We now solve the homogenous form of Eq.~(\ref{Apdyn}) by writing
\begin{equation}
\label{Appsi}
\tilde{\psi}_k=a_1(t) e^{\mu_1 t} + a_2(t) e^{\mu_2 t},
\end{equation}
where
\begin{eqnarray}
\mu_1&=& -\alpha/2 + \sqrt{\alpha^2/4+\hat{L}_k}, \\
\mu_2&=& -\alpha/2 - \sqrt{\alpha^2/4+\hat{L}_k},
\end{eqnarray}
and where $a_1$ and $a_2$ are functions of time and restricted to satisfy
\begin{equation}
\label{Aprest}
\frac{d a_1}{dt} e^{\mu_1 t} + \frac{d a_2}{dt}  e^{\mu_2 t}=0.
\end{equation}
Combining Eq.~(\ref{Appsi}) and Eq.~(\ref{Apdyn}) we derive the relation
\begin{equation}
\label{Apnhm}
\mu_1e^{\mu_1 t}\frac{d a_1}{dt}+\mu_2e^{\mu_2 t}\frac{d a_2}{dt} =\tilde{f}_k.
\end{equation}
Additionally combining Eq.~(\ref{Apnhm}) and Eq.~(\ref{Aprest}) and integrating over one time step we obtain
\begin{eqnarray}
a_1&=&\frac{1}{\mu_1-\mu_2} \int_0^{\Delta t} e^{-\mu_1 t'} \tilde{f}_k(t') dt' + a_{10},\\
a_2&=&\frac{-1}{\mu_1-\mu_2} \int_0^{\Delta t} e^{-\mu_2 t'} \tilde{f}_k(t') dt' + a_{20},
\end{eqnarray}
where $a_{10}$ and $a_{20}$ are integration constants.
Now combining everything
\begin{eqnarray}
\tilde{\psi}_k(t+\Delta t) &=& \frac{1}{\mu_1-\mu_2}\left[e^{\mu_1(t+\Delta t)}\int_t^{t+\Delta t} e^{-\mu_1 t'}\tilde{f}_k(t')dt'
\right.\nonumber\\
&&\left. -e^{\mu_2(t+\Delta t)}\int_t^{t+\Delta t} e^{-\mu_2 t'}\tilde{f}_k(t')dt'\right]\nonumber\\
&&+a_{10}e^{\mu_1(t+\Delta t)}+a_{20}e^{\mu_2 (t+\Delta t)},\\
\tilde{u}_k(t+\Delta t) &=& \frac{1}{\mu_1-\mu_2}\left[\mu_1e^{\mu_1(t+\Delta t)}\int_t^{t+\Delta t} e^{-\mu_1 t'}\tilde{f}_k(t')dt'\right.\nonumber\\
&&\left.-\mu_2e^{\mu_2(t+\Delta t)}\int_t^{t+\Delta t} e^{-\mu_2 t'}\tilde{f}_k(t')dt'\right]\nonumber\\
&&+\mu_1a_{10}e^{\mu_1(t+\Delta t)}+\mu_2a_{20}e^{\mu_2(t+\Delta t)}.\nonumber\\
\end{eqnarray}
We integrate over the range $t=0$ to $t=\Delta t$ and require that at time $t=0$
\begin{eqnarray}
\tilde{\psi}_k(t=0)= \tilde{\psi}_{k0} &=& a_{10} +a_{20},\\
\tilde{u}_k(t=0)=\tilde{u}_{k0} &=& \mu_1 a_{10} + \mu_2 a_{20}.
\end{eqnarray}
Solving for the integration constants $a_{10}$ and $a_{20}$
\begin{eqnarray}
a_{10} &=& \frac{(\tilde{u}_{k0}-\mu_2\tilde{\psi}_{k0})}{\mu_1-\mu_2},\\
a_{20} &=& \frac{-(\tilde{u}_{k0}-\mu_1\tilde{\psi}_{k0})}{\mu_1-\mu_2}.
\end{eqnarray}
To calculate the integral terms we use the approach presented in the Appendix of Ref.~\onlinecite{Mellenthinetal2008} by expanding $\tilde{f}_k$ around $t'=t$
\begin{widetext}
\begin{equation}
\frac{e^{\mu_1(t+\Delta t)}}{\mu_1-\mu_2} \int_t^{t+\Delta t} e^{-\mu_1 t'}\tilde{f}_k(t')dt'=
\frac{e^{\mu_1(t+\Delta t)}}{\mu_1-\mu_2} \int_t^{t+\Delta t}dt'e^{-\mu_1 t'}\times
\left(\tilde{f}_k(t)+\frac{\tilde{f}_k(t)-\tilde{f}_k(t-\Delta t)}{\Delta t}(t'-t)\right).
\end{equation}
Integrating the first term in the integral 
\begin{equation}
\frac{e^{\mu_1(t+\Delta t)}}{\mu_1-\mu_2} \int_t^{t+\Delta t} dt'e^{-\mu_1 t'}\tilde{f}_k(t)= %\nonumber\\
\frac{-e^{-\mu_1(t+\Delta t)} +e^{-\mu_1 t}}{\mu_1(\mu_1-\mu_2)}\tilde{f}_k e^{\mu_1(t+\Delta t)}= %\nonumber\\
\frac{(e^{\mu_1\Delta t}-1)}{\mu_1(\mu_1-\mu_2)}\tilde{f}_k(t).
\end{equation}
Integrating the second term in the integral we see
\begin{equation}
\frac{e^{\mu_1(t+\Delta t)}}{\mu_1-\mu_2} \int_t^{t+\Delta t} dt'e^{-\mu_1 t'}\frac{\tilde{f}_k(t)-\tilde{f}_k(t-\Delta t)}{\Delta t}(t'-t)= %\nonumber\\
\frac{(\tilde{f}_k(t)-\tilde{f}_k(t-\Delta t))}{\mu_1(\mu_1-\mu_2)}\left( \frac{e^{\mu_1\Delta t}}{\mu_1 \Delta t} - \frac{1}{\mu_1\Delta t}-1\right).
\end{equation}
The full dynamics then take on the form
\begin{eqnarray}
\label{Appsifull}
\tilde{\psi}_k(t+\Delta t) &=& \frac{1}{\mu_1-\mu_2}\left( -(\tilde{\psi}_k(t)\mu_2-\tilde{u}_k(t))e^{\mu_1\Delta t}+(\tilde{\psi}_k(t)\mu_1-\tilde{u}_k(t))e^{\mu_2\Delta t}\right)\nonumber\\
&+&\frac{\tilde{f}_k(t)}{\mu_1(\mu_1-\mu_2)}(e^{\mu_1\Delta t }-1)+\frac{(\tilde{f}_k(t)-\tilde{f}_k(t-\Delta t))}{\mu_1(\mu_1-\mu_2)}\left(
\frac{e^{\mu_1 \Delta t}}{\mu_1 \Delta t}-\frac{1}{\mu_1 \Delta t}-1\right)\nonumber \\
&-&\frac{\tilde{f}_k(t)}{\mu_2(\mu_1-\mu_2)}(e^{\mu_2\Delta t }-1)-\frac{(\tilde{f}_k(t)-\tilde{f}_k(t-\Delta t))}{\mu_2(\mu_1-\mu_2)}\left(
\frac{e^{\mu_2 \Delta t}}{\mu_2 \Delta t}-\frac{1}{\mu_2 \Delta t}-1\right),\\
\label{Apufull}
\tilde{u}_k(t+\Delta t) &=& \frac{1}{\mu_1-\mu_2}\left( -\mu_1(\tilde{\psi}_k(t)\mu_2-\tilde{u}_k(t))e^{\mu_1\Delta t}+\mu_2(\tilde{\psi}_k(t)\mu_1-\tilde{u}_k(t))e^{\mu_2\Delta t}\right)\nonumber\\
&+&\frac{\tilde{f}_k(t)}{(\mu_1-\mu_2)}(e^{\mu_1\Delta t }-1)+\frac{(\tilde{f}_k(t)-\tilde{f}_k(t-\Delta t))}{(\mu_1-\mu_2)}\left(
\frac{e^{\mu_1 \Delta t}}{\mu_1 \Delta t}-\frac{1}{\mu_1 \Delta t}-1\right)\nonumber \\
&-&\frac{\tilde{f}_k(t)}{(\mu_1-\mu_2)}(e^{\mu_2\Delta t }-1)-\frac{(\tilde{f}_k(t)-\tilde{f}_k(t-\Delta t))}{(\mu_1-\mu_2)}\left(
\frac{e^{\mu_2 \Delta t}}{\mu_2 \Delta t}-\frac{1}{\mu_2 \Delta t}-1\right).
\end{eqnarray}
%\hrule
\end{widetext}
The non-conserved dynamics take on a similar form
\begin{equation}
\partial_{tt}\psi +\alpha\partial_t\psi=-\frac{\delta F}{\delta \psi}+\mu,
\end{equation}
where $\mu$ is an externally imposed chemical potential.  The only changes to the dynamics are in the operator $\hat{L}_k$ and the non linear term $\tilde{f}_k$
\begin{eqnarray}
\hat{L}_k&=&-k^{4}+2k^2+(\epsilon-1),\\
\tilde{f}_k &=& \int d\vec{r} e^{i\vec{k}\cdot \vec{r}} (-\psi^3+\mu).
\end{eqnarray}

%%%%%%%%%%%%%%%%%%%%%%%%%%%%%%%%%%%%%%%%%%%%%%%%%%%%%%%%%%%%
\section{Liquid layer width for PFC with noise}
\label{wnoise}

\begin{figure}
\centering
\includegraphics[width=0.5\textwidth, angle=0]{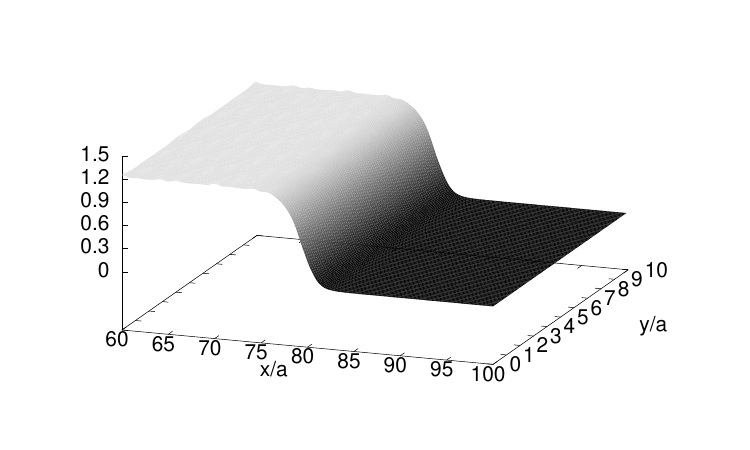}
\caption{Surface plot of the order parameter $\tilde{\rho}(\vec{r})$ defined in Eq.~(\ref{gauss}) through a solid-liquid interface for PFC without noise. The graph shows the order parameter plotted for a single slice in the x-y plane (for a single value of z).  In the solid we see $\tilde{\rho}_s\approx 1.22$ while in the liquid $\tilde{\rho}_l= 0$, with a $\tanh$ like interpolation between the two values. $a$ is one lattice spacing.
}
\label{transsl}
\end{figure}

\begin{figure}
\centering
\includegraphics[width=0.5\textwidth, angle=0]{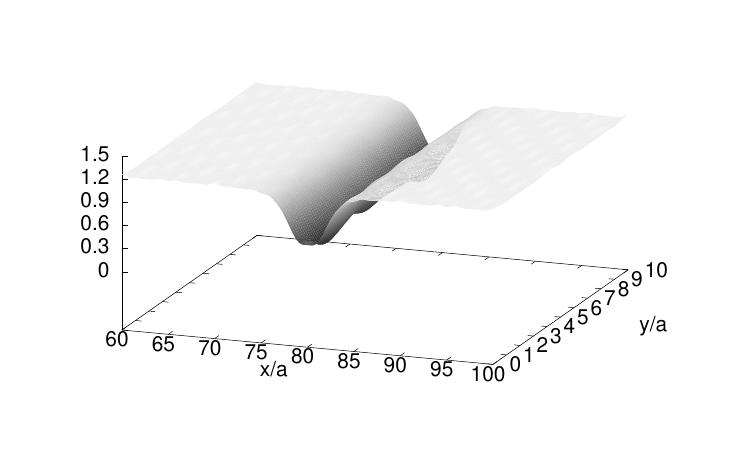}
\caption{Surface plot of the order parameter $\tilde{\rho}(\vec{r})$ defined in Eq.~(\ref{gauss}) through a grain-boundary interface for PFC without noise. The graph shows the order parameter plotted for a single slice in the x-y plane (for a single value of z).  In the solid we see $\tilde{\rho}_s\approx 1.22$, however the order parameter does not take on its minimum value of $\tilde{\rho}=0$ between the two crystals. This is consistent with an interface where the two grains are not well separated. $a$ is one lattice spacing.
}
\label{transdeterm}
\end{figure}

The inclusion of noise in the PFC model demands an alternative definition of the melt layer thickness $W$.  The method of calculating the liquid layer width for PFC without noise by using the excess mass no longer works once we have added noise.  This is due to the relatively large instantaneous fluctuations in the amount of mass in the system.  Instead we introduce a wavelet transform with methodology similar to a  transform used to measure the local orientation of a two-dimensional PFC hexagonal crystal \cite{Singer06}. In our case we extract an orientation independent field which differentiates continuously between solid and liquid. We define a convolution of the square of the gradient of the density field $\psi$ with a normalized Gaussian function,
\begin{equation}
\tilde{\rho}(\vec{r})=\int G(|\vec{r}-\vec{r}'|) |\nabla \psi(\vec{r}')|^2d\vec{r}',
\end{equation}
where
\begin{equation}
G=\frac{1}{\sqrt{\pi\sigma_2}}e^{-(x^2+y^2+z^2)/2\sigma_2},\label{gauss}
\end{equation}
and $\sigma_2=2.5a$ where $a$ is one lattice spacing.
The convolution returns an orientation independent field whose amplitude depends on how close to a perfect solid the crystal is.  As seen in Fig.~\ref{transsl} this results in a smooth transition from the solid to the liquid through a solid-liquid interface in the deterministic case. The behavior of the order parameter is different for a grain-boundary interface in the deterministic case, as seen in Fig.~\ref{transdeterm}. $\tilde{\rho}$ never reaches its minimum value of $0$ in between the two solid regions. This behavior is consistent with an interface where the two grains are not well separated.
\begin{figure}
\centering
\includegraphics[width=0.5\textwidth, angle=0]{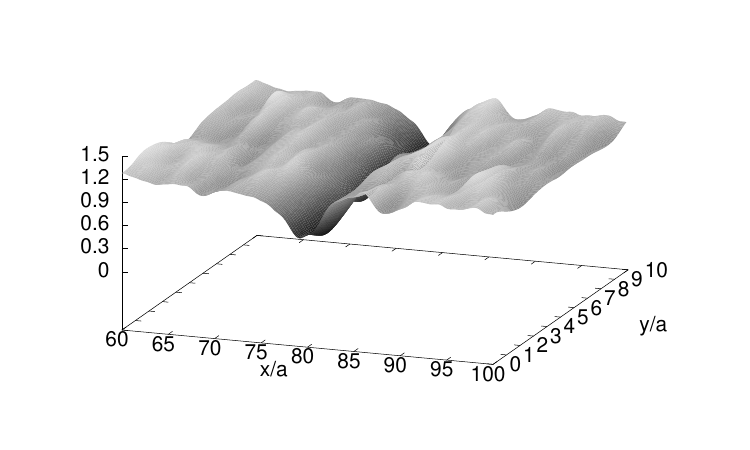}
\caption{Surface plot of the order parameter $\tilde{\rho}(\vec{r})$ defined in Eq.~(\ref{gauss}) through a grain-boundary interface for PFC with noise. The graph shows the order parameter plotted for a single slice in the x-y plane (for a single value of z). As compared to the deterministic case, there is more variation in the order parameter in the solid and the central liquid like region. This makes it necessary to average over several points to identify the liquid and solid values of the order parameter, as detailed in the text. $a$ is one lattice spacing.
}
\label{transnoise}
\end{figure}

The behavior of the order parameter when including noise is more complicated, as seen in Fig.~\ref{transnoise}. Instead of attempting to determine the liquid layer width $W$ in one calculation, we separate the order parameter into an array of functions $\tilde{\rho}(x,y_0,z_0)$, where $x$ takes on its full range of values and $y_0$ and $z_0$ refer to a particular point in the $y-z$ plane. We then fit the following equation

\begin{eqnarray}
\tilde{\rho}_l(y_0,z_0)+\frac{\tilde{\rho}_s(y_0,z_0)+\tilde{\rho}_l(y_0,z_0)}{2}(1-\tanh(b_l(x-x_l)))\nonumber\\
+\frac{\tilde{\rho}_s(y_0,z_0)-\tilde{\rho}_l(y_0,z_0)}{2}(1+\tanh(b_r(x-x_r))),\nonumber\\
\end{eqnarray}
to $\tilde{\rho}(x,y_0,z_0)$ for each combination of $y_0$ and $z_0$. $b_l,b_r,x_l, $ and $ x_r$ are fit parameters while $\tilde{\rho}_s(y_0,z_0)$ is the value of the order parameter in the solid and is obtained by averaging $\tilde{\rho}(x,y_0,z_0)$ over $5$ different values of $x$, where $x$ is several lattice spacings away from the interface. $\tilde{\rho}_l(y_0,z_0)$ is defined as the average of the two smallest values of $\tilde{\rho}(x,y_0,z_0)$.

$W(y_0,z_0)$, the liquid layer width for each set $y_0,z_0$, is defined as $x_r-x_l$. The total liquid layer width $W$ is then the average of $W(y_0,z_0)$ over all $y_0,z_0$.

%%%%%%%%%%%%%%%%%%%%%%%%%%%%%%%%%%%%%%%%%%%%%%
\section{Linking PFC and DFT}
\label{appenDFT}

To obtain the value of the only adjustable parameter, $\epsilon$, we derive an expression to relate $\epsilon$ to material parameters taken from MD simulations, namely the peak and curvature at the peak of the liquid structure factor. While already derived in Ref.~\onlinecite{WuKarma2007} we use method presented in Ref.~\onlinecite{Wuetal2010} for the two-mode PFC model that more directly relates these quantities to classical DFT \cite{RamYus79,HaymetOxtoby1981,Lairdetal1987,Singh1991,HarwellOxtoby1984,ShenOxtoby1996a,ShenOxtoby1996b}. We vary $\psi$ around its liquid value $\psi = \bar{\psi}_l+\delta\psi$ , where $\bar{\psi}_l$ is the density in the liquid, as determined from the phase coexistence conditions between solid and liquid, and substitute into Eq.~(\ref{ffunc}). 
We see that (dropping terms of $\delta\psi$ higher than quadratic order)
\begin{equation}
{\Delta\cal{F}} = \frac{\lambda q_0^4}{g}\int d\vec{r} \left[ \frac{\delta\psi}{2}[a+3\bar{\psi}_l^2\lambda q_0^4+\lambda(\nabla^2+q_0^2)^2]\delta\psi\right].
\label{delf}
\end{equation}
Defining the Fourier transform,
\begin{equation}
\delta\psi=\int\frac{d\vec{k}}{(2\pi)^{3/2}}\delta\psi_k e^{i\vec{k}\cdot\vec{r}},
\end{equation}
and substituting this definition into Eq. (\ref{delf}),
\begin{eqnarray}
{\Delta\cal{F}} =&& \frac{\lambda q_0^4}{g}\int\int \frac{d\vec{k}d\vec{k}'}{{(2\pi)}^3}\frac{\delta\psi_k\delta\psi_{k'}}{2} \Big[(a+3\bar{\psi}^2_l \lambda q_0^4 +\lambda(-k^2  \nonumber\\
&&+q_0^2)^2)
\int d\vec{r}e^{i(\vec{k}+\vec{k}')\cdot \vec{r}} \Big]\nonumber\\
=&&\frac{\lambda q_0^4}{g(2\pi)^{3/2}} \int d\vec{k}\frac{\delta\psi_k \delta\psi_{-k}}{2} \left[a+3\bar{\psi}^2_l\lambda q_0^4\right. \nonumber\\
&&\left.+\lambda(-k^2+q_0^2)^2\right].
\end{eqnarray}
We can compare this equation to a standard formulation from classical DFT,
\begin{eqnarray}
{\Delta\cal{F}} _{DFT}&=&\frac{k_B T}{2}\int\int d\vec{r}d\vec{r}'\nonumber\\
&& \delta n(\vec{r})\left[ \frac{\delta(\vec{r}-\vec{r}')}{n_0}-C(|\vec{r}-\vec{r}'|)\right]\delta n(\vec{r}'),\nonumber\\
\end{eqnarray}
where the density variations $\delta n(\vec{r})$ are related to the PFC order parameter via Eq.~(\ref{pfc::eq1})
\begin{equation}
\delta n(\vec{r})=n(\vec{r})-n_0=\delta \phi(\vec{r})=\sqrt{\frac{\lambda q_0^4}{g}}\delta\psi(\vec{r}),
\end{equation}
and
\begin{equation}
C(k)=n_0\int d\vec{r} C(|\vec{r}|) e^{-i\vec{k}\cdot \vec{r}}
\end{equation}
is the Fourier transform of the direct correlation function. 
Substituting in the Fourier transforms as before
\begin{equation}
{\Delta\cal{F}} _{DFT}=\frac{\lambda q_0^4}{g}\frac{k_B T}{2n_0(2\pi)^{3/2}}\int d\vec{k}\delta \psi_k\delta\psi_{-k}[1-C(k)].
\end{equation}
Equating the two free-energy densities and using the expression for the liquid structure factor $S(k)=1/[1-C(k)] $ we see that
\begin{equation}
S(k)=\frac{k_B T}{n_0 \{a+3\lambda q_0^4\bar{\psi}_l^2+\lambda(-k^2+q_0^2)^2\}}.
\end{equation}
Evaluated at the peak of the structure factor ($k=q_0$) we see,
\begin{equation}
a+3\lambda q_0^4\bar{\psi}_l^2=\frac{k_BT}{n_0 S(q_0)}.
\label{eqna}
\end{equation}

The multiscale expansion using $\epsilon^{1/2}$ as small parameter is discussed in \cite{Wuetal2006} in detail, and we repeat only the results here.
To this end the density field is $\psi$ in expanded in powers of $\epsilon^{1/2}$,
\begin{equation}
\psi(\vec{r})=\psi_0(\vec{r})\epsilon^{1/2}+\psi_1(\vec{r})\epsilon+\psi_2(\vec{r})\epsilon^{3/2}+\cdot\cdot\cdot,
\end{equation}
and we can expand the average solid and liquid densities in the same way
\begin{eqnarray}
\bar{\psi}_s&=&\psi_{s0}\epsilon^{1/2}+\psi_{s1}\epsilon+\psi_{s2}\epsilon^{3/2}+\cdot\cdot\cdot,\\
\bar{\psi}_l&=&\psi_{l0}\epsilon^{1/2}+\psi_{l1}\epsilon+\psi_{l2}\epsilon^{3/2}+\cdot\cdot\cdot.
\end{eqnarray}
From the common tangent construction one obtains at the different powers of $\epsilon$
\begin{eqnarray}
\psi_{s0}&=&\psi_{l0}\equiv\psi_c=-\sqrt{\frac{45}{103}},\\
\psi_{s1}&=&\psi_{l1}=0.
\end{eqnarray}
We see that in the small $\epsilon$ limit the size of the solid-liquid coexistence region $\Delta \bar{\psi}=\bar{\psi}_s-\bar{\psi}_l\approx(\psi_{s2}-\psi_{l2})\epsilon^{3/2}$ is much smaller than the average density which scales like $\epsilon^{1/2}$ justifying our dropping of higher order terms in the small $\epsilon$ regime.

Using Eq.~(\ref{epsconv}) to substitute for $a$ and the small $\epsilon$ justification for dropping higher order terms to define $\bar{\psi}_l=\psi_c\epsilon^{1/2}$ we get,
\begin{equation}
\epsilon=\frac{-k_BT}{n_0 S(q_0)\lambda q_0^4(1-3\psi_c^2)}.
\end{equation}
Taking the second derivative of $C(k)$ and evaluating at $k=q_0$ we obtain a value for $\lambda$,
\begin{equation}
\lambda=-\frac{k_BT C''(q_0)}{8q_0^2n_0},
\label{eqnlambda}
\end{equation}
see also Eq.~(\ref{gl::eq1}), and we thus get
\begin{equation}
\epsilon=\frac{8}{(1-3\psi_c^2)q_0^2S(q_0)C''(q_0)},
\end{equation}
which matches an expression originally derived in \cite{WuKarma2007}.  Using the values for $\psi_c$, $S(q_0)$ and $C''(q_0)$ reported in \cite{WuKarma2007} (see also Table \ref{table3}) we get $\epsilon=0.0923$.

In order to convert the dimensionless PFC results back into dimensional units we must also define $a$, $\lambda$ and $g$. 
$a$ and $\lambda$ can be defined through Eqs.~(\ref{eqna}) and~(\ref{eqnlambda}) respectively. 
To calculate $g$ we first recognize that
\begin{equation}
n_0u_i=\sqrt{\frac{\lambda q_0^4}{g}}\epsilon^{1/2} A_i^0,
\end{equation}
where $u_i$ are the dimensional amplitudes (as opposed to the dimensionless amplitudes $A_i$) and $A_i^0 =  \epsilon^{-1/2}A_s$. 
Substituting $u_i=u_s$, which can be obtained from MD simulations \cite{Wuetal2006} we obtain, 

\begin{equation}
g=\lambda q_0^4 \epsilon \left( -\frac{2}{15}\psi_c +\frac{1}{15}\sqrt{5-11\psi_c^2}\right)^2/(n_0^2u_s^2).
\end{equation}
This is a correction from the expression first presented in \cite{WuKarma2007}.

 \end{document}